\newcommand\apjcls{1}
\newcommand\aastexcls{2}
\newcommand\othercls{3}

\newcommand\papercls{\aastexcls}
\documentclass[tighten, times, twocolumn]{aastex62}  

\if\papercls \apjcls
\usepackage{apjfonts}
\else\if\papercls \othercls
\usepackage{epsfig}
\usepackage{margin}
\usepackage{times}

\fi\fi
\usepackage{ifthen}
\usepackage{natbib}
\usepackage{bm}
\usepackage{amssymb, amsmath}
\usepackage{appendix}
\usepackage{etoolbox}
\usepackage[T1]{fontenc}
\usepackage{paralist}
\usepackage{newtxtext,newtxmath}
\if\papercls \apjcls
\newcommand\aas{\ref@jnl{AAS Meeting Abstracts}}
\newcommand\dps{\ref@jnl{AAS/DPS Meeting Abstracts}}
\newcommand\maps{\ref@jnl{MAPS}}
\else\if\papercls \othercls
\usepackage{astjnlabbrev-jh}
\fi\fi
\usepackage{xcolor}

\if\papercls \aastexcls
\hypersetup{citecolor=blue, 
            linkcolor=blue, 
            menucolor=blue, 
            urlcolor=blue}  
\else
\usepackage[
bookmarks=true,           
bookmarksnumbered=true,   
colorlinks=true,          
citecolor=blue,           
linkcolor=blue,           
menucolor=blue,           
urlcolor=blue,            
linkbordercolor={0 0 1},  
pdfborder={0 0 1},
frenchlinks=true]{hyperref}
\fi
\usepackage{float}

\if\papercls \othercls

\else

\fi

\providecommand{\adsurl}[1]{\href{#1}{ADS}}

\makeatletter
\patchcmd{\NAT@citex}
  {\@citea\NAT@hyper@{%
     \NAT@nmfmt{\NAT@nm}%
     \hyper@natlinkbreak{\NAT@aysep\NAT@spacechar}{\@citeb\@extra@b@citeb}%
     \NAT@date}}
  {\@citea\NAT@nmfmt{\NAT@nm}%
   \NAT@aysep\NAT@spacechar\NAT@hyper@{\NAT@date}}{}{}

\patchcmd{\NAT@citex}
  {\@citea\NAT@hyper@{%
     \NAT@nmfmt{\NAT@nm}%
     \hyper@natlinkbreak{\NAT@spacechar\NAT@@open\if*#1*\else#1\NAT@spacechar\fi}%
       {\@citeb\@extra@b@citeb}%
     \NAT@date}}
  {\@citea\NAT@nmfmt{\NAT@nm}%
   \NAT@spacechar\NAT@@open\if*#1*\else#1\NAT@spacechar\fi\NAT@hyper@{\NAT@date}}
  {}{}
\makeatother

\makeatletter
\DeclareRobustCommand{\lowcase}[1]{\@lowcase#1\@nil}
\def\@lowcase#1\@nil{\if\relax#1\relax\else\MakeLowercase{#1}\fi}
\makeatother

\DeclareSymbolFont{UPM}{U}{eur}{m}{n}
\DeclareMathSymbol{\umu}{0}{UPM}{"16}
\let\oldumu=\umu
\renewcommand\umu{\ifmmode\oldumu\else\math{\oldumu}\fi}

\if\papercls \othercls

\else

\fi

\let\oldsim=\sim
\renewcommand\sim{\ifmmode\oldsim\else\math{\oldsim}\fi}
\let\oldpm=\pm
\renewcommand\pm{\ifmmode\oldpm\else\math{\oldpm}\fi}
\newcommand\by{\ifmmode\times\else\math{\times}\fi}

\newbox{\wdbox}
\renewcommand\c{\setbox\wdbox=\hbox{,}\hspace{\wd\wdbox}}
\renewcommand\i{\setbox\wdbox=\hbox{i}\hspace{\wd\wdbox}}

\newcount\timect
\newcount\hourct
\newcount\minct
\newcommand\now{\timect=\time \divide\timect by 60
         \hourct=\timect \multiply\hourct by 60
         \minct=\time \advance\minct by -\hourct
         \number\timect:\ifnum \minct < 10 0\fi\number\minct}

\newcommand{\Ra}{\mathrm{Ra_T}}

\catcode`@=11

\newcommand\comment[1]{}

\newcommand\commenton{\catcode`\%=14}

\renewcommand\math[1]{$#1$}
\newcommand\mathshifton{\catcode`\$=3}

\let\atab=&
\newcommand\atabon{\catcode`\&=4}

\let\oldmsp=\sp
\let\oldmsb=\sb
\def\sp#1{\ifmmode
           \oldmsp{#1}%
         \else\strut\raise.85ex\hbox{\scriptsize #1}\fi}
\def\sb#1{\ifmmode
           \oldmsb{#1}%
         \else\strut\raise-.54ex\hbox{\scriptsize #1}\fi}
\newbox\@sp
\newbox\@sb
\def\sbp#1#2{\ifmmode%
           \oldmsb{#1}\oldmsp{#2}%
         \else
           \setbox\@sb=\hbox{\sb{#1}}%
           \setbox\@sp=\hbox{\sp{#2}}%
           \rlap{\copy\@sb}\copy\@sp
           \ifdim \wd\@sb >\wd\@sp
             \hskip -\wd\@sp \hskip \wd\@sb
           \fi
        \fi}
\def\msp#1{\ifmmode
           \oldmsp{#1}
         \else \math{\oldmsp{#1}}\fi}
\def\msb#1{\ifmmode
           \oldmsb{#1}
         \else \math{\oldmsb{#1}}\fi}

\def\supon{\catcode`\^=7}

\def\subon{\catcode`\_=8}

\def\supsubon{\supon \subon}

\newcommand\actcharon{\catcode`\~=13}

\newcommand\paramon{\catcode`\#=6}

\comment{And now to turn us totally on and off...}

\newcommand\reservedcharson{ \commenton  \mathshifton  \atabon  \supsubon 
                             \actcharon  \paramon}

\catcode`@=12
\reservedcharson

\if\papercls \apjcls

\else

\fi

\shorttitle{Heat transport by compositionally-driven convection}

\shortauthors{Fuentes \emph{et al.}}

\begin{document}

\title{Heat transport and convective velocities in compositionally-driven convection in neutron star and white dwarf interiors}


\author{J.~R.~Fuentes}
\affiliation{\rm Department of Applied Mathematics, University of Colorado Boulder, Boulder, CO 80309-0526, USA}
\affiliation{\rm Department of Physics and Trottier Space Institute, McGill University, Montreal, QC H3A 2T8, Canada}

\author{Andrew Cumming}
\affiliation{\rm Department of Physics and Trottier Space Institute, McGill University, Montreal, QC H3A 2T8, Canada}

\author{Matias Castro-Tapia}
\affiliation{\rm Department of Physics and Trottier Space Institute, McGill University, Montreal, QC H3A 2T8, Canada}

\author{Evan H. Anders}
\affiliation{\rm Center for Interdisciplinary Exploration and Research in Astrophysics, Northwestern University, Evanston, Illinois 60201, USA
}

\email{jofu5477@colorado.edu}

\begin{abstract}
We investigate heat transport associated with compositionally-driven convection driven by crystallization at the ocean-crust interface in accreting neutron stars, or growth of the solid core in cooling white dwarfs. We study the effect of thermal diffusion and rapid rotation on the convective heat transport, using both mixing length theory and numerical simulations of Boussinesq convection. We determine the heat flux, composition gradient and P\'eclet number, $\mathrm{Pe}$ (the ratio of thermal diffusion time to convective turnover time) as a function of the composition flux.
We find two regimes of convection with a rapid transition between them as the composition flux increases. 
At small Pe, the ratio between the heat flux and composition flux is independent of Pe, because the loss of heat from convecting fluid elements due to thermal diffusion is offset by the smaller composition gradient needed to overcome the reduced thermal buoyancy.
At large Pe, the temperature gradient approaches the adiabatic gradient, saturating the heat flux.
We discuss the implications for neutron star and white dwarf cooling. Convection in neutron stars spans both regimes. We find rapid mixing of neutron star oceans, with a convective turnover time of order weeks to minutes depending on rotation. Except during the early stages of core crystallization, white dwarf convection is in the thermal-diffusion-dominated fingering regime. We find convective velocities much smaller than recent estimates for crystallization-driven dynamos. The small fraction of energy carried as kinetic energy calls into question the effectiveness of crystallization-driven dynamos as an explanation for observed white dwarf magnetic fields.
\end{abstract}

\keywords{convection -- stars:neutron -- stars: white dwarfs -- X-rays: binaries}

\section{Introduction}

When a multicomponent plasma freezes, the composition of the solid is typically different from the composition of the liquid. If the solid preferentially retains heavy elements, the liquid left behind is lighter and buoyant, driving convection. The compositionally-driven convection transports light elements outwards and mixes the liquid region. This process has been studied in the context of dense interiors of white dwarfs \citep{Stevenson1980,Mochkovitch1983,Isern1997} and accreting neutron stars \citep{MC11,MC14,MC15}, and also occurs in planetary interiors, e.g.~Earth \citep{Fearn1981}, the Moon \citep{Laneuville2014,Scheinberg2015} and Mercury \citep{Manglik2010}. Depending on the phase diagram, another possibility is that heavy elements preferentially go into the liquid phase, so that solid crystals float upwards.
This distillation process has recently been suggested to be occurring in white dwarfs, driven by chemical separation of $^{22}$Ne between the liquid and solid phases \citep{Blouin2021} (see also \citealt{Mochkovitch1983}).

Redistribution of elements in white dwarf interiors is important because the gravitational energy released can prolong white dwarf cooling. The large increase in the number of white dwarfs with well-determined distances from Gaia \citep{GentileFusillo2021} has enabled the cooling delay associated with crystallization to be definitely detected. The slowed cooling is visible as an increased density of white dwarfs in the HR diagram or luminosity function \citep{Tremblay2019}. One puzzling feature in the HR diagram known as the Q-branch indicates an additional cooling delay in a small fraction of massive white dwarfs \citep{Cheng2019}. Explanations for the delay have focused on the redistribution of elements ($^{22}$Ne in particular) within the white dwarf \citep{Bauer2020,Blouin2021,Camisassa2021,Caplan2020}.

Neutron stars in low mass X-ray binaries accrete enough mass over their lifetimes to replace the entire neutron star crust (eg.~\citealt{Suleiman2022}). The accreted light elements first undergo thermonuclear burning in the surface layers, generating a complex mixture of heavy elements that forms a liquid ocean \citep{Bildsten1995}. At the base of the ocean, compressed matter continuously freezes and forms solid crust as accretion continues \citep{Brown1998}. In sources that undergo transient accretion outbursts, the neutron star cools in quiescence and the liquid ocean refreezes (see \citealt{Wijnands2017} for a review of transiently accreting neutron stars). \cite{Horowitz2007} showed that chemical separation between liquid and solid phases is expected for the mixtures found in neutron star oceans, with lighter elements typically left behind in the liquid phase (see \citealt{Mckinven2016} and \citealt{Caplan2018} for a survey of different compositions). \cite{MC11,MC14,MC15} studied the compositional changes and heat transport in the ocean in these different scenarios.

Unlike in many planetary interiors, where convection is driven by both compositional and thermal buoyancy, crystallization-driven convection in white dwarf and neutron star interiors occurs in a part of the star that is thermally-stable to convection, i.e.~has a sub-adiabatic temperature gradient. This is because of the large thermal conductivity from degenerate electrons which can transport the cooling luminosity and latent heat of crystallization with only a small temperature gradient. In this case, when compositionally-driven convection occurs in a thermally-stratified background, convection transports heat in the opposite direction to the composition flux \citep{Loper1978,MC11}. Rising fluid elements adiabatically expand and cool down to a temperature that is lower than their surroundings. This cools the surroundings, giving an effective heat flow that is directed downwards. 
By transporting heat towards the liquid/solid interface, convection acts in a similar way to the latent heat.
\cite{MC14} showed that this
changes the cooling rate of neutron stars following accretion outbursts, an observable signature of the newly-forming crust and its composition.

When calculating the convective heat flux in neutron star oceans, \cite{MC11,MC15} assumed that the fluid motions would be adiabatic. However, a large enough thermal conductivity could cause rising parcels of fluid to lose a significant amount of energy by thermal diffusion, reducing the effective heat flux. The likely importance of thermal diffusion in white dwarf convection was pointed out by \cite{Stevenson1980} and included in estimates of the convective velocities by \cite{Mochkovitch1983} and \cite{Isern1997}. The P\'eclet number, the ratio of thermal diffusion time to convective turnover time was estimated to be $\approx 0.3$ by \cite{Isern1997}, implying that this effect is important. The transport of heat by compositionally-driven convection does not appear to have been considered in white dwarfs; instead, it is usually assumed that the liquid region above the crystallization front mixes rapidly, and the resulting change in energy is put directly into the model as a localized heat source \citep{Isern1997,Isern2000}.

Interest in compositionally-driven convection in white dwarfs has also been recently revived with the suggestion of \cite{Isern2017} that it leads to a magnetic dynamo in crystallizing white dwarfs \citep{Schreiber2021b,Schreiber2021a,Belloni2021,Camisassa2022,Ginzburg2022,Schreiber2022}. Using the scaling of \cite{Christensen2009} for a saturated dynamo, \cite{Isern2017} found that fields up to $\sim 1\ \mathrm{MG}$ could be generated. However, whether the dynamo is in the saturated regime depends on the convective turnover time, and estimates of the convective velocity differ significantly. \cite{Isern2017} found $v_c\approx 30\ {\rm km\ s^{-1}}$ by considering rising carbon-enriched liquid bubbles released at the crystallization front, whereas \cite{Ginzburg2022} argued that the velocity should be much lower,  $\sim 100\ \mathrm{cm\ s^{-1}}$, based on the available convective energy flux. Both of these velocity estimates are significantly larger than previous estimates for (non-magnetic) compositionally-driven convection. \cite{Mochkovitch1983} found $v_c\sim 10^{-6}\ {\rm cm\ s^{-1}}$ for non-rotating or $\sim 0.1\ {\rm cm\ s^{-1}}$ for rapidly-rotating white dwarfs.

In this paper, we revisit compositionally-driven convection in dense stellar interiors. Our goal is to determine the expected convective velocities and convective heat flux for accreting neutron stars and cooling white dwarfs. We apply stellar mixing length theory to the case of compositionally-driven convection, and use numerical simulations to demonstrate that heat is indeed transported inwards and test the mixing length theory predictions. The mixing length theory is presented in section 2, where we derive expressions for the heat flux and convective velocities, and discuss the steady-state outcome in which the inwards heat flux due to convection is balanced by an outwards conductive heat flux. In section 3, we present our numerical simulations of Boussinesq convection in the non-rotating case and compare with mixing length theory. We conclude in section 4 with a discussion of how our results apply to white dwarfs and neutron stars.

\section{Mixing length theory for compositionally-driven convection} \label{sec:MLT}

In this section, we use mixing length theory to investigate the size of the heat flux associated with compositionally-driven convection, and the expected convective velocities. We first write down mixing length theory including thermal diffusion (\S\ref{sec:leakage}), and then discuss the expected heat flux (\S\ref{sec:heatflux}) and convective velocities in the non-rotating and rapidly-rotating limits (\S\ref{sec:rotation}).

We then investigate the steady-state in which the inwards convective flux is balanced by outwards conduction (\S\ref{sec:P\'eclet}).

\subsection{Mixing length theory including thermal diffusion}\label{sec:leakage}

In mixing length theory, the heat and composition fluxes are written in terms of the excess temperature $DT$ or composition $DX$ carried by a fluid element, $F_H =\rho v_c c_P DT$ and $F_X = \rho v_c  DX$, where $v_c$ is the convective velocity, $c_P$ is the specific heat capacity at constant pressure, and $\rho$ the density.
For simplicity, we assume a mixture of two elements, so that the composition can be described by only one variable, here chosen to be $X$, the mass fraction of the lighter component\footnote{The results can be easily generalized to more complex mixtures, e.g.~\cite{MC15}. We also make the approximation that the excess entropy carried by fluid elements is $DS\approx c_P DT/T$, ignoring any contribution to the entropy from compositional differences. Again, this can be included in a straightforward way, writing the heat flux as $\rho v_c T DS$, but is typically a small correction \citep{MC15}.}.

We include the effect of thermal diffusion following the formulation of mixing length theory discussed by \cite{KippenhahnWiegert}, which is based on \cite{BohmVitense1958} (see also \citealt{Henyey1965} and \citealt{Gough1977}). The temperature excess is written as 
\begin{equation}
	DT = \left(\nabla-\nabla_e\right)T{\ell \over 2H_P},
\end{equation}
where $\nabla = \left.d\ln T/d\ln P\right|_\star$ is the temperature gradient in the star, $\nabla_e$ is the rate of change of temperature with pressure experienced by the fluid element, $\ell$ is the mixing length, and $H_P$ the pressure scale height. Similarly, we can write $DX/X = \nabla_X (\ell/2H_P)$, where $\nabla_X = \left.d\ln X/d\ln P\right|_\star$ is the composition gradient in the star. 

The heat and composition fluxes are then given by 
\begin{equation}\label{eq:heatflux}
F_H = \rho v_c c_P DT = \rho v_c c_P T\left(\nabla-\nabla_e\right){\ell \over 2H_P},
\end{equation}

and
\begin{equation}\label{eq:FX}
F_X = \rho v_c X\nabla_X{\ell \over 2H_P}.
\end{equation}
The sign of these fluxes is such that a positive flux is in the upwards direction. For example, an outwards flux of light elements is associated with a gradient $\nabla_X>0$, i.e. the mass fraction of light elements increases with pressure. Note that the composition flux gives the mass of light elements crossing unit area per unit time (i.e.~in cgs the units of $F_X$ are $\mathrm{g\ cm^{-2}\ s^{-1}}$). In this paper, we consider situations in which $\nabla_{\mathrm{ad}}>\nabla_e>\nabla>0$, so that the system is stable against thermal convection and $DT<0$.

By considering the exchange of energy by thermal diffusion with the surroundings as the fluid element moves, \cite{KippenhahnWiegert} derive an expression for $\nabla_e$
\begin{equation}\label{eq:Pe}
	{\nabla_e-\nabla_\mathrm{ad}\over \nabla-\nabla_e} = 
 {9\over 2}{K\over \rho c_P\ell v_c} = {9\over 2}{\kappa_T\over \ell v_c} \equiv {9\over 2}{1\over \mathrm{Pe}}\, ,
\end{equation}
where $\kappa_T$ is the thermal diffusivity and we define the dimensionless P\'eclet number $\mathrm{Pe}\equiv \ell v_c/\kappa_T$. The numerical prefactor of $9/2$ in equation \eqref{eq:Pe} depends on assumptions about the shape of the fluid element and the temperature distribution (see discussion in \citealt{Henyey1965}). For example, \cite{Mihalas} following \cite{BohmVitense1958} give a prefactor of 3 instead, whereas \cite{Henyey1965} have a prefactor of $2\pi^2\approx 20$. Here, instead of adopting any particular value, we keep in mind that it is model-dependent and treat it as a free parameter $C$. Replacing the $9/2$ by $C$ in equation \eqref{eq:Pe} gives
\begin{equation}\label{eq:nne}
\nabla-\nabla_e = \left({{\rm Pe}\over {C + {\rm Pe}}}\right)\left(\nabla-\nabla_{\rm ad}\right).
\end{equation}
When the convective motions are rapid, $\mathrm{Pe}\gg 1$ and $\nabla_e\rightarrow \nabla_\mathrm{ad}$ as expected since the motions become adiabatic. In the opposite limit in which the convective motions are slow and thermal diffusion can act, $\mathrm{Pe}\ll 1$ and $\nabla_e\rightarrow \nabla$, so that the fluid element is able to adjust its temperature to follow the background temperature gradient.

\subsection{The heat flux in compositionally-driven convection}
\label{sec:heatflux}

Taking the ratio of equations \eqref{eq:heatflux} and \eqref{eq:FX}, the convective velocity and mixing length drop out, giving the heat flux in terms of the composition flux, 
\begin{equation}\label{eq:FHc}
	{F_H\over F_X} = -{c_PT\over X}{\nabla_e-\nabla\over \nabla_X}.
\end{equation}
This shows that transport of composition is associated also with a transport of heat, provided the fluid elements experience a different temperature evolution with pressure compared to the background. Equation \eqref{eq:nne} shows that $\nabla_e$ ranges from $\nabla$ to $\nabla_\mathrm{ad}$ as $\mathrm{Pe}$ goes from small to large values. In a background that is stably-stratified thermally, ie. with $\nabla_\mathrm{ad}>\nabla$, this means that $\nabla_e\geq\nabla$, giving a heat flux oppositely-directed to the composition flux.

The fact that $\nabla_e$ approaches $\nabla$ for $\mathrm{Pe}\ll 1$ (eq.~[\ref{eq:nne}]) acts to reduce the heat flux. However, the composition gradient in the convection zone also depends on $\mathrm{Pe}$, since the effective thermal stratification, $\nabla - \nabla_e$, is reduced at low $\mathrm{Pe}$ when thermal diffusion is efficient. This means that a smaller composition gradient is needed to maintain the convective motions. To see this, consider the typical density contrast in the convection zone,
\begin{equation}
{D\rho\over\rho}\approx -{\chi_T\over\chi_\rho} {DT\over T} -{\chi_X\over\chi_\rho} {DX\over X},  
\end{equation} 
 where $\chi_T = \left.\partial\ln P/\partial\ln T\right|_{\rho, X}$, $\chi_\rho = \left.\partial\ln P/\partial\ln \rho\right|_{T, X}$, and $\chi_X = \left.\partial\ln P/\partial\ln X\right|_{\rho, T}$. The density contrast determines the buoyant acceleration $\propto -D\rho/\rho$. Written in terms of the gradients, 

\begin{eqnarray}\label{eq:Drhograd}
{D\rho\over\rho}&\approx& -{\ell\over 2H_P}\left[{\chi_T\over\chi_\rho}(\nabla-\nabla_e) +{\chi_X\over\chi_\rho} \nabla_X\right]\nonumber\\
&\approx& -{\ell\over 2H_P}{\chi_X\over \chi_\rho}\left[\nabla_X-\nabla_{X,\mathrm{crit}}\right],
\end{eqnarray} 
where we define the critical composition gradient
\begin{equation}\label{eq:DXcrit}
    \nabla_{X,\mathrm{crit}} = {\chi_T\over \chi_X}\left(\nabla_e-\nabla\right) = {\chi_T\over \chi_X}\left(\nabla_\mathrm{ad}-\nabla\right) \left({\mathrm{Pe}\over C+\mathrm{Pe}}\right).
\end{equation}
For adiabatic displacements (large $\mathrm{Pe}$), where $\nabla_e\rightarrow\nabla_\mathrm{ad}$, $D\rho<0$ in equation \eqref{eq:Drhograd} is equivalent to the Ledoux criterion for convection, $\chi_T\left(\nabla-\nabla_\mathrm{ad}\right)+\chi_X\nabla_X>0$, and so $\nabla_{X,\mathrm{crit}}$ in this limit is the composition gradient needed to be unstable to convection according to the Ledoux criterion. At small $\mathrm{Pe}$, thermal diffusion lowers the effective thermal stratification, reducing $\nabla_{X,\mathrm{crit}}$, and allowing convection to occur for smaller composition gradients. This is the regime of fingering or thermohaline convection\footnote{In the limit $\mathrm{Pe}\ll 1$ and assuming $\nabla_X\approx\nabla_{X,\mathrm{crit}}$, equation~\eqref{eq:DXcrit} agrees with the prescription for convection in the MESA code \citep{Paxton2013} based on \cite{Ulrich1972} and \cite{Kippenhahn1980}. To see this, write the diffusion coefficient in eq.~(14) of \cite{Paxton2013} as $D_\mathrm{th}=v_c\ell$, in which case their expression reduces to the small $\mathrm{Pe}$ limit of eq.~\eqref{eq:DXcrit}. The efficiency parameter for thermohaline convection $\alpha_\mathrm{th}$ is related to our shape parameter by $\alpha_\mathrm{th}=2C/3$.}. A similar expression to equation~\eqref{eq:Drhograd} was previously written down by \cite{Mochkovitch1983} for the case $C=1$.

If the convection is efficient in the sense that $\nabla_X - \nabla_{X,\mathrm{crit}} \ll \nabla_{X,\mathrm{crit}}$\footnote{This is analagous to the efficient regime of thermal convection where $\nabla-\nabla_\mathrm{ad}\ll \nabla_\mathrm{ad}$. It is interesting to note that whereas loss of energy by thermal diffusion causes thermal convection to become less efficient, here we find that thermal diffusion makes compositionally-driven convection more efficient because it reduces $\nabla_{X,\mathrm{crit}}$, allowing for efficient composition transport with $\nabla_X-\nabla_{X,\mathrm{crit}}\ll \nabla_{X,\mathrm{crit}}$.} (as is the case in our problem at Pe $\lesssim 1$, see Sect.~\ref{sec:P\'eclet} and Fig.~\ref{fig:analyticPe}), the reduction in $\nabla_e - \nabla$ at small Pe is exactly offset by the reduction in $\nabla_X$, so the ratio $F_H/F_X$ is actually independent of Pe. To see this more explicitly, we can write the heat flux in terms of $\nabla_{X,\mathrm{crit}}$, giving
\begin{equation}\label{eq:FHFXPe}
	{F_H\over F_X} = -{c_PT\over X}{\chi_X\over \chi_T}\left({\nabla_{X,\mathrm{crit}}\over \nabla_X}\right).
\end{equation}
This relation between $F_H$ and $F_X$ is the same as derived by \cite{MC11} under the assumption that fluid elements move adiabatically (the only difference is that $\nabla_{X,\mathrm{crit}}$ in that case is given by the large $\mathrm{Pe}$ limit of eq.~[\ref{eq:DXcrit}]).

\subsection{Convective velocity and effect of rotation}
\label{sec:rotation}

We can estimate the extent to which $\nabla_X$ exceeds $\nabla_{X,\mathrm{crit}}$ by writing the expression for the convective velocity
\begin{equation}\label{eq:vc2}
v_c^2 \approx {g\ell\over 4}{D\rho\over\rho}\approx {g\ell^2\over 8H_P} {\chi_X\over \chi_\rho} \left(\nabla_X-\nabla_{X,\mathrm{crit}}\right),
\end{equation}
where we take the numerical prefactors $1/4$ and $1/8$ from the particular formulation of mixing length theory we are using \citep{KippenhahnWiegert}.
Using the definition $\mathrm{Pe}=\ell v_c/\kappa_T$ and defining a Rayleigh number 
\begin{equation}
    \mathrm{Ra_T} = {gH_P^3\chi_T\nabla_\mathrm{ad}\over\chi_\rho\kappa_T^2},
\end{equation}
we obtain
\begin{equation}\label{eq:XminusXcrit}
    {\chi_X\over\chi_T\nabla_\mathrm{ad}}(\nabla_X-\nabla_{X,\mathrm{crit}}) = {8\over \mathrm{Ra_T}}\left({H_P\over\ell}\right)^4 \mathrm{Pe}^2.
\end{equation}

For the large $\mathrm{Ra_T}$ in astrophysical applications (e.g. see Sects.~\ref{sec:implications_NS} and \ref{sec:implications_WD}), the term on the right hand side will be small as long as $\mathrm{Pe}$ is not too large, so that taking $\nabla_X\approx \nabla_{X,\mathrm{crit}}$ should be a good approximation. However, for a large enough composition flux, this term can become important as we will see below.

Equation \eqref{eq:vc2} assumes that the velocity of fluid elements is set by the buoyant acceleration acting over a mixing length. In rapidly-rotating convection, Coriolis forces modify the force balance and change the convective velocity. We estimate the effect of rapid rotation following the scaling relations of \cite{Aurnou2020}, who considered the balance between Coriolis, inertial and buoyancy terms in rapidly-rotating convection (CIA balance). Simulations of non-magnetic rapidly-rotating convection in planetary cores give support to this scaling \citep{Guervilly2019}. Rewriting equation (24) of \cite{Aurnou2020} in our notation, this balance can be expressed as 
\begin{equation}\label{eq:CIA}
    {v_c^2\over L^2}\sim {2\Omega v_c\over H_P}\sim {g\over H_P}{\chi_X\over \chi_\rho}\left(\nabla_X-\nabla_{X,\mathrm{crit}}\right).
\end{equation}
We assume that the lengthscale associated with convective motions in the direction of the rotation vector is the pressure scale height $H_P$, while $L$ is the lengthscale associated with motions perpendicular to the rotation vector. The first and last terms of equation~\eqref{eq:CIA} give an expression for the convective velocity that has the same functional form as equation~\eqref{eq:vc2} but with the replacement $\ell\rightarrow L$. The first and second terms in equation~\eqref{eq:CIA} give the ratio between perpendicular and parallel scales as
\begin{equation}
    {L\over H_P} \approx \left({v_c\over 2\Omega H_P}\right)^{1/2}\approx \mathrm{Ro}^{1/2},
\end{equation}
where we define the Rossby number $\mathrm{Ro}\equiv v_c/2\Omega H_P$. 

These scalings suggest that we can estimate the effect of rapid rotation on the convective velocity by making the substitution $\ell\rightarrow L\approx \mathrm{Ro}^{1/2}H_P$ in the non-rotating result. The Rossby number is given in terms of $\mathrm{Pe}$ (which is now defined as $\mathrm{Pe}\equiv v_cL/\kappa_T$) by 
\begin{equation}\label{eq:Ro}
\mathrm{Ro}\equiv {v_c\over 2\Omega H_P} = \mathrm{Pe}^{2/3}\mathrm{Ta}^{-1/3},
\end{equation}
where we define the Taylor number
\begin{equation}
\mathrm{Ta}\equiv {4\Omega^2 H_P^4\over\kappa_T^2}.
\end{equation}
With these scalings, we find
\begin{equation}\label{eq:XminusXcritrot}
    {\chi_X\over\chi_T\nabla_\mathrm{ad}}(\nabla_X-\nabla_{X,\mathrm{crit}}) = {8\over \mathrm{Ra_T}}{\mathrm{Pe}^2\over \mathrm{Ro}^2}={8\over \mathrm{Ra_T}}\mathrm{Ta}^{2/3}\mathrm{Pe}^{2/3}.
\end{equation}
Comparing to equation \eqref{eq:XminusXcrit}, we see that rapid rotation ($\mathrm{Ro}\ll 1)$ acts to steepen the composition gradient. Even so, the large value of $\mathrm{Ra_T}$ in astrophysical scenarios means that $\nabla_X$ will remain very close to $\nabla_{X,\mathrm{crit}}$ in many cases.

\subsection{The steady-state balance with thermal conduction}

\label{sec:P\'eclet}

We now consider the consequences of the mixing length theory outlined above in a situation with a specified outwards flux of light elements $F_X$.  As the compositionally-driven convection transports heat inwards, the temperature gradient will steepen until the outwards conductive heat flux balances the inwards convective heat flux\footnote{The outwards conductive flux and inwards convective flux will not exactly cancel. For example, in a cooling white dwarf there must be a net outwards cooling luminosity. In neutron star envelopes, \cite{MC15} considered a steady-state in which the net heat flux was inwards, carrying nuclear energy released in a low density H/He burning shell into the neutron star interior. In both cases, the latent heat needs to be removed from the crystallization front. For simplicity here we assume that any net flux is small compared to the convective heat flux. In addition, for very strongly-driven convection, the kinetic energy flux can become significant and should be added to equation \eqref{eq:energy_balance}. However, the kinetic energy flux is small for the composition fluxes expected in compact objects (see Appendix \ref{app:KE}) and so we do not include it here.},
\begin{equation}\label{eq:energy_balance}
   \rho c_P\kappa_T {T\nabla \over H_P} = - \rho v_c c_P T\left(\nabla-\nabla_e\right){\ell \over 2H_P}.
\end{equation}

Solving for the steady-state temperature gradient gives $\nabla = \nabla_e \mathrm{Pe}/(2 + \mathrm{Pe})$, or using equation \eqref{eq:nne},
\begin{equation}\label{eq:nablass}
    \nabla = \nabla_\mathrm{ad} {\mathrm{Pe}^2\over\mathrm{Pe}^2 + 2\mathrm{Pe} + 2C}.
\end{equation}
When $\mathrm{Pe}\ll 1$, conduction acts efficiently on the timescale of convection, so that a small temperature gradient $\nabla\approx \nabla_\mathrm{ad}\mathrm{Pe}^2/2C$ is sufficient for conduction to balance the convective heat flux. However, when convection is driven very strongly and the convective velocities become large,  $\mathrm{Pe}\gg 1$, the steady-state temperature gradient approaches the adiabatic gradient ($\nabla\rightarrow\nabla_{\rm ad}$), 
so that the convective flux ($\propto \nabla_{\rm ad}-\nabla$) saturates to a point where it can be balanced by the conductive flux along the adiabat\footnote{In reality, the temperature gradient may saturate below $\nabla_\mathrm{ad}$. Once the temperature gradient reaches the slope of the liquidus curve $\nabla_L\approx 1/4 < \nabla_{\rm ad}\approx 1/3$, large portions of liquid will freeze, shutting down convection. 
\cite{MC14,MC15} found that the system then becomes time-dependent with periodic freezing and melting of large regions near the liquid/solid boundary, on average maintaining a gradient $\nabla\approx\nabla_L$. For simplicity, we ignore this effect in this section.}. 

We can write an expression for $\mathrm{Pe}$ in terms of the composition flux using equation \eqref{eq:FX}, which gives the convective velocity
$v_c \approx (F_X/ \rho X \nabla_X)(2H_P/\ell)$, or
\begin{equation}\label{eq:evalPe}
    \mathrm{Pe} = {v_c \ell\over \kappa_T} \approx \left({H_P^2\over \kappa_T}\right) \left({2F_X\over \rho H_P X}\right)\nabla_X^{-1}.
\end{equation}
The first term is the thermal diffusion time across the pressure scale height $t_\mathrm{therm}=H_P^2/\kappa_T$. The second term is related to the timescale on which the light elements are being injected into the layer. For example, consider a region of a star with mass $\Delta M\sim 4\pi r^2\rho H_P$. If light elements are being injected at a rate $\dot M_X = \Delta M \dot X$, the composition flux is $F_X\sim \Delta M \dot X/4\pi r^2 = \rho H_P \dot X$, and the second term in equation \eqref{eq:evalPe} is $F_X/\rho H_P X\sim \dot{X}/X$. We therefore define the timescale $t_X=\rho H_P X/F_X$, giving\footnote{A similar expression for $\mathrm{Pe}$ was previously obtained by \cite{Mochkovitch1983} (their eq.~[23]) and \cite{Isern1997} (their eq.~[32]) for the case where $\nabla_X=\nabla_{X,\mathrm{crit}}$ and assuming $C=1$.}
\begin{equation}\label{eq:Pe_estimate}
	\mathrm{Pe}\approx {t_{\rm therm}\over t_X}{2\over \nabla_X}.
\end{equation}
 We see that the P\'eclet number is set by both the ratio of thermal and injection timescales and the composition gradient. For fixed timescales, a smaller composition gradient requires a larger velocity to transport the composition.

 \begin{figure}
   \centering
    \includegraphics[width=1\columnwidth]{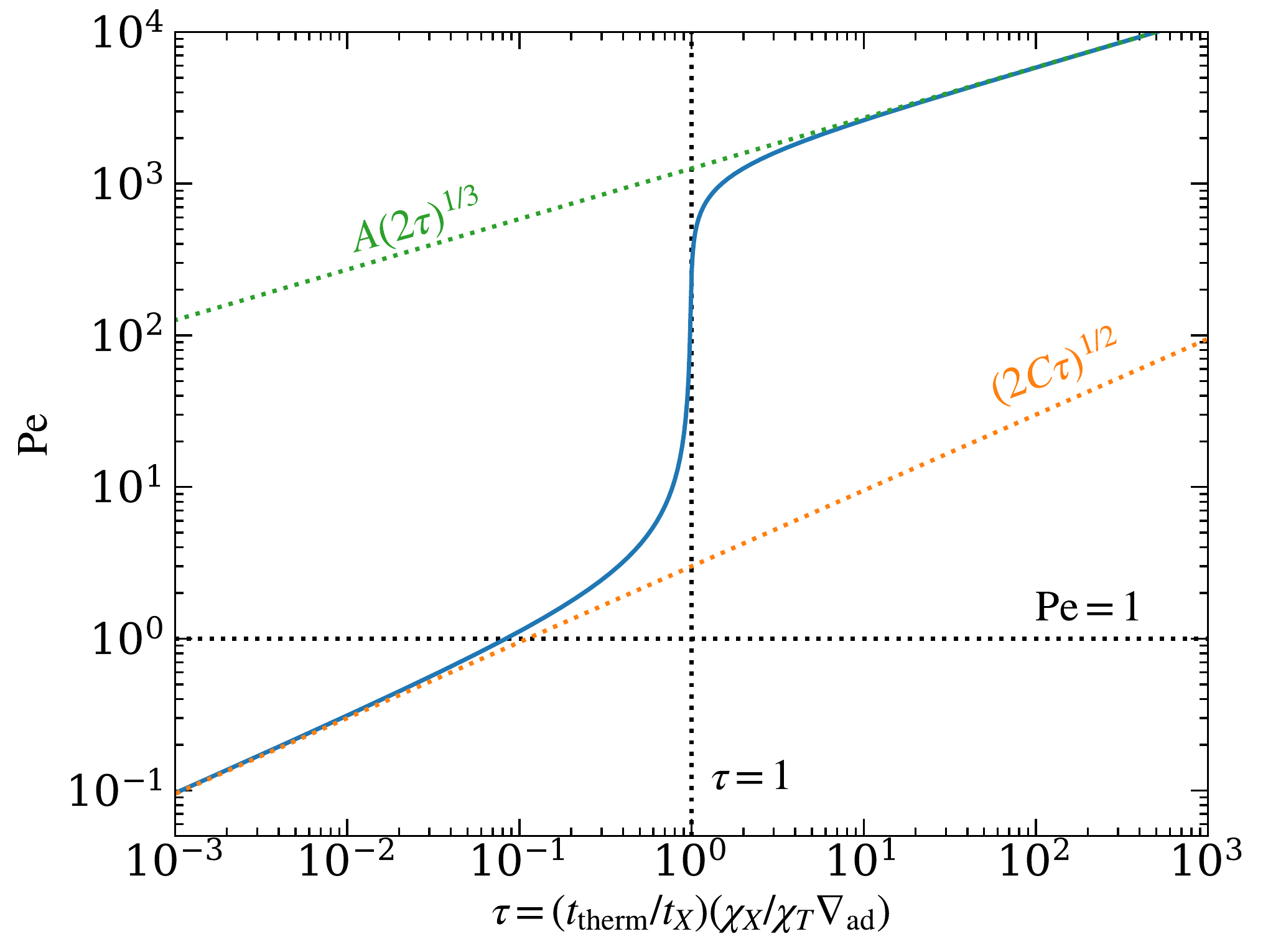}
    \includegraphics[width=1\columnwidth]{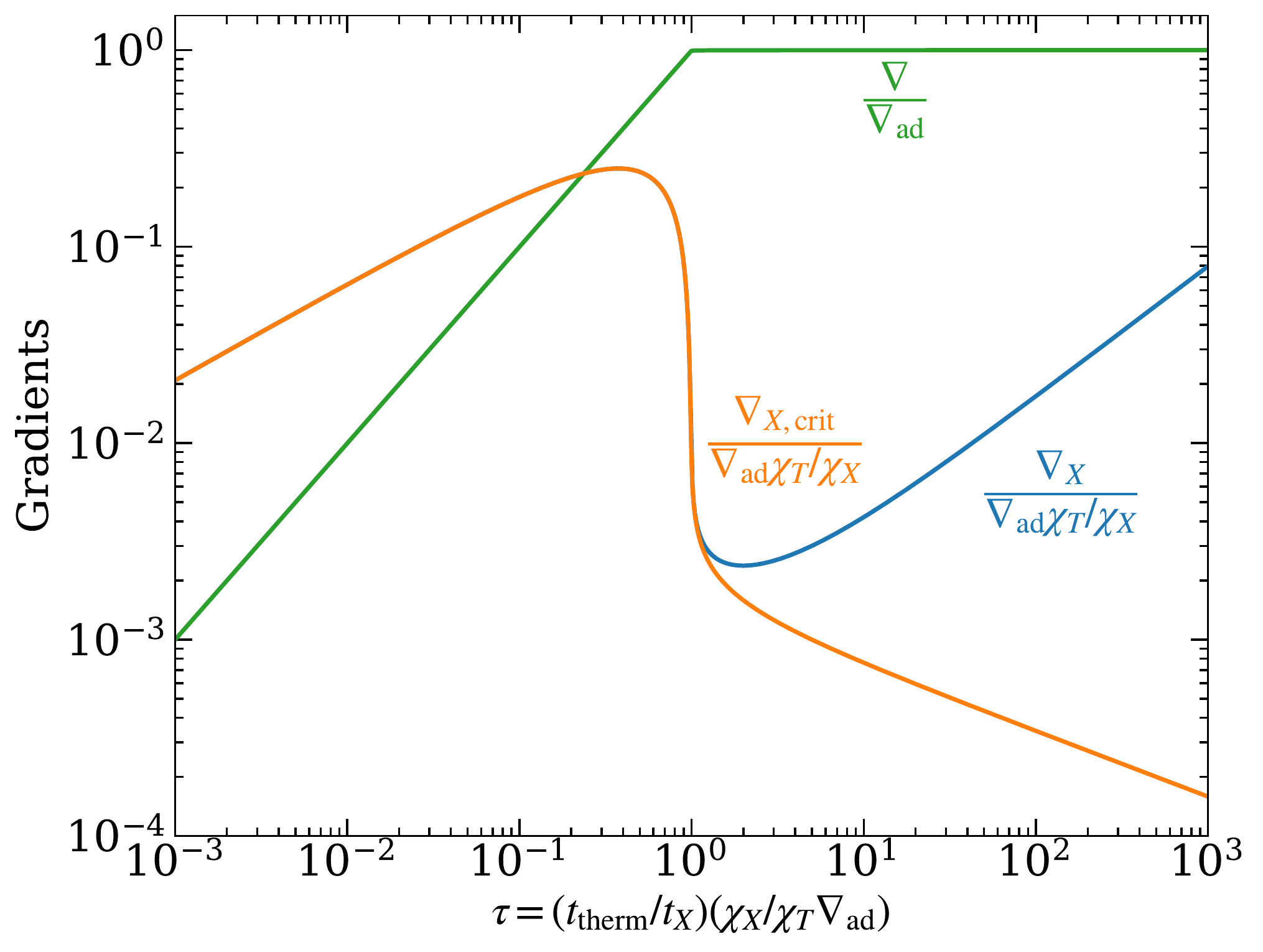}
    \caption{The steady-state P\'eclet number ({\em top panel}) and temperature and composition gradients ({\em bottom panel}) as a function of the driving parameter $\tau\propto F_X$ (eq.~\ref{eq:tau}). Near $\tau=1$, $\mathrm{Pe}$ rapidly transitions from the small $\tau$ solution of eq.~\eqref{eq:smallbranch} (lower dotted line) to the large $\tau$ solution given by eq.~\eqref{eq:largebranch} (upper dotted line). For this example, we set $C=9/2$ and $A=10^3$ ($\Ra\sim 10^{10}$).}
    \label{fig:analyticPe}
\end{figure}

Equations \eqref{eq:nablass}, and \eqref{eq:Pe_estimate} both relate $\mathrm{Pe}$ to one of the gradients $\nabla$ or $\nabla_X$. Adding a third relation, either equation \eqref{eq:XminusXcrit} for no rotation or equation \eqref{eq:XminusXcritrot} for rapid rotation, we can solve for $\mathrm{Pe}$, $\nabla$ and $\nabla_X$. Before presenting the solution, it is useful to define the dimensionless parameter 
\begin{equation}\label{eq:tau}
     \tau = \left({t_{\rm therm}\over t_X}\right)\left({\chi_X\over \chi_T\nabla_\mathrm{ad}}\right).
\end{equation}
which is a measure of the composition flux driving convection. This can also be written explicitly in terms of $F_X$ as
\begin{equation}\label{eq:tau2}
     \tau = {F_X\over F_{H,\mathrm{ad}}} \left({c_PT\over X}{\chi_X\over\chi_T}\right),
\end{equation}
where $F_{H,\mathrm{ad}}=\rho c_P \kappa_T T\nabla_\mathrm{ad}/H_P$ is the heat flux conducted along the thermal adiabat. Comparing with equation \eqref{eq:FHFXPe} we see that $\tau$ is a measure of the effect of the convection on the temperature gradient: when $\tau=1$, the value of $F_X$ is such that the associated convective heat flux for efficient convection ($\nabla_X\approx \nabla_{X,\mathrm{crit}}$) is equal to $F_{H,\mathrm{ad}}$. This means that for $\tau\ll 1$, the heat flux can be balanced by a small temperature gradient $\nabla=\tau\nabla_\mathrm{ad}$. The temperature gradient is much shallower than the adiabat, thermal diffusion is efficient, and $\mathrm{Pe}$ is small. For $\tau\gg 1$, the heat flux for efficient convection exceeds $F_{H,\mathrm{ad}}$, the composition gradient steepens $\nabla_X>\nabla_{X,\mathrm{crit}}$ to reduce the heat flux to $\approx F_{H,\mathrm{ad}}$, the temperature gradient is close to the adiabat ($\nabla\approx\nabla_\mathrm{ad}$) with inefficient thermal diffusion and large $\mathrm{Pe}$.

The full solution for the non-rotating case can be written as 
\begin{equation}\label{eq:solution_nonrot}
    \tau = {\mathrm{Pe}^2\over \mathrm{Pe}^2 + 2\mathrm{Pe} + 2C} + {1\over 2} \left({\mathrm{Pe}\over A}\right)^3,
\end{equation}
where
\begin{equation}\label{eq:Adef}
   A = {1\over 2} \Ra^{1/3}
\left({\ell\over H_P}\right)^{4/3}.   
\end{equation}
Equation \eqref{eq:solution_nonrot} gives $\mathrm{Pe}(\tau)$ which can then be used to obtain the gradients $\nabla$ and $\nabla_X$ using equations \eqref{eq:nablass} and \eqref{eq:Pe_estimate} respectively. In particular, the composition gradient is given by $\nabla_X = (\chi_T\nabla_\mathrm{ad}/\chi_X) (2\tau  / \mathrm{Pe})$.

 \begin{figure}
   \centering
    \includegraphics[width=1\columnwidth]{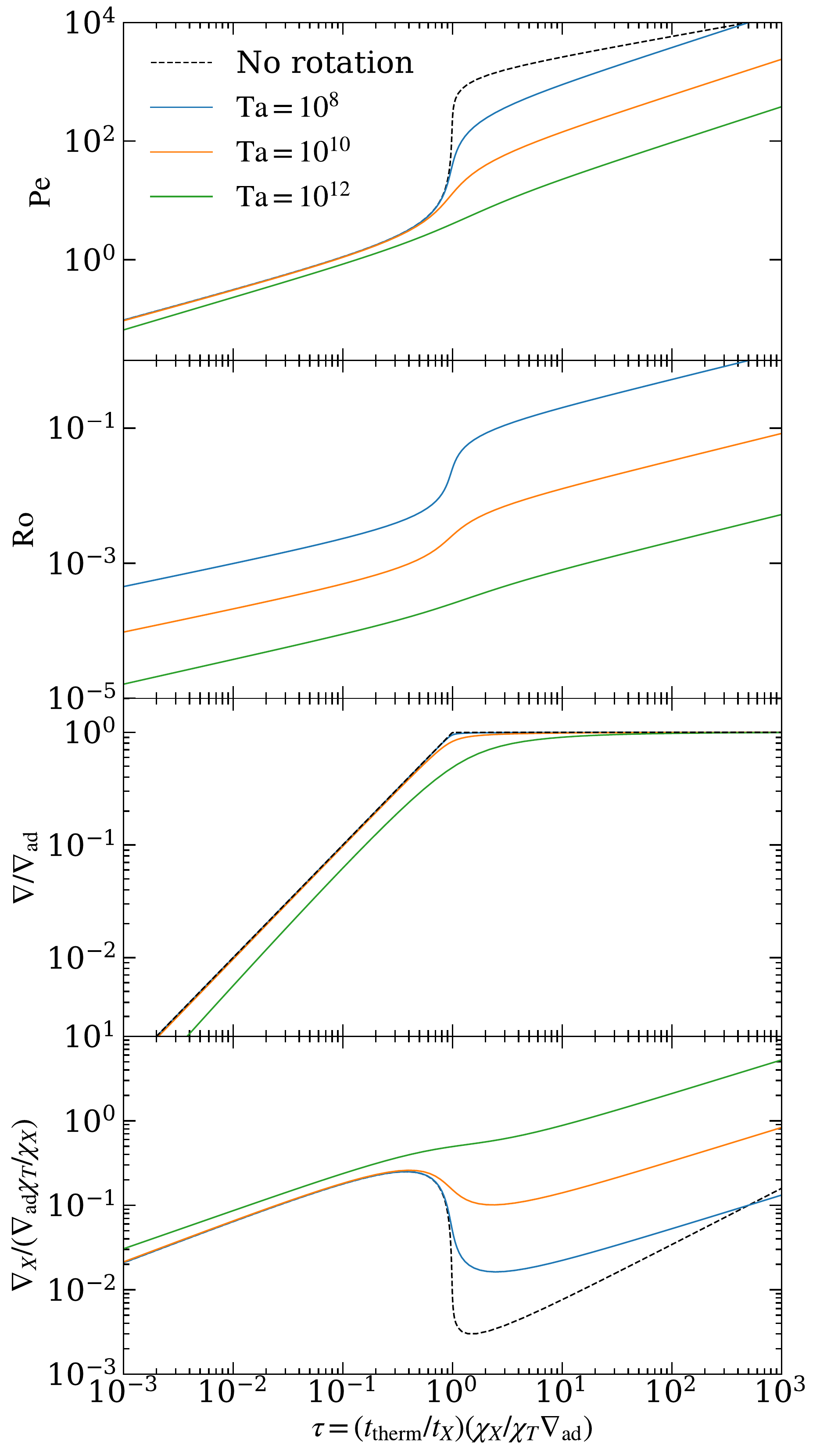}
    \caption{The effect of rapid rotation on the steady-state solutions. We show the non-rotating solution from Fig.~\ref{fig:analyticPe} as the dashed black line. The other curves show the effect of increasing rotation on this model, with $\mathrm{Ta}=10^8, 10^{10}$, and $10^{12}$. Rapid rotation has only a small effect on the temperature gradient/heat flux, but leads to smaller convective velocities and larger composition gradients.}
    \label{fig:analyticPe_rotation}
\end{figure}

An example for particular choices of $A$ and $C$ is shown in Figure~\ref{fig:analyticPe}.
The solution for $\mathrm{Pe}(\tau)$ (top panel) has two branches: at small $\tau$, $\nabla_X\approx \nabla_{X,\mathrm{crit}}\propto\mathrm{Pe}$ and $\nabla\ll \nabla_\mathrm{ad}$, so that equation \eqref{eq:Pe_estimate} gives
\begin{equation}\label{eq:smallbranch}
	\mathrm{Pe}\approx (2C\tau)^{1/2} \hspace{1cm}(\tau\ \mathrm{small}),
\end{equation}
while at large $\tau$, the composition gradient is set by the right hand term in equation \eqref{eq:XminusXcrit}, giving
\begin{equation}\label{eq:largebranch}
    \mathrm{Pe}\approx A(2\tau)^{1/3}, \hspace{1cm}(\tau\ \mathrm{large}).
\end{equation}
The value of $\mathrm{Pe}$ makes a rapid transition between these two branches at $\tau=1$.
The lower panel of Figure \ref{fig:analyticPe} shows the gradients. The temperature gradient closely follows $\nabla=\nabla_\mathrm{ad} \tau$ for $\tau<1$ and $\nabla=\nabla_\mathrm{ad}$ for $\tau>1$. The composition gradient shows a more complicated behaviour. For $\tau<1$, it is very close to $\nabla_X=\nabla_{X,\mathrm{crit}}$. At small values of $\tau$, this gives $\nabla_X$ increasing with $\tau$, $\nabla_X\approx (\nabla_\mathrm{ad}\chi_T/\chi_X)(2\tau/C)^{1/2}$. As $\tau\rightarrow 1$, $\nabla\rightarrow\nabla_\mathrm{ad}$, decreasing the thermal buoyancy and therefore $\nabla_{X,\mathrm{crit}}$, which leads to the rapid decrease in $\nabla_X$ near $\tau=1$ in Figure \ref{fig:analyticPe}. For $\tau>1$, $\nabla_X$ increases with $\tau$ again as it starts to significantly exceed $\nabla_{X,\mathrm{crit}}$. For large $\tau$, $\nabla_X\approx (\nabla_\mathrm{ad}\chi_T/\chi_X)(2\tau)^{2/3}/A$.

For rapid rotation, we use equation \eqref{eq:XminusXcritrot} instead of \eqref{eq:XminusXcrit}. The solution is 
\begin{equation}\label{eq:tau_eq_with_omega}
    \tau = {\mathrm{Pe}^2\over \mathrm{Pe}^2 + 2\mathrm{Pe} + 2C} +\left({\mathrm{Ta}^{2/3}\over 2 A^3}\right)\mathrm{Pe}^{5/3}.
\end{equation}
An example is shown in Figure \ref{fig:analyticPe_rotation} which shows the effect of increasing rotation on the model from Figure \ref{fig:analyticPe}.
As long as $\mathrm{Ta}^{2/3}\lesssim A^3$ (corresponding approximately to $\mathrm{Ta}^{2/3}\lesssim \mathrm{Ra_T}$), then the first term in equation \eqref{eq:tau_eq_with_omega} dominates for $\tau<1$. The results for $\mathrm{Pe}$ and the gradients are therefore the same as without rotation\footnote{In the limit of very rapid rotation, when $\mathrm{Ta}^{2/3}>A^3$, the last term in equation \eqref{eq:tau_eq_with_omega} dominates for all $\tau$, giving $\mathrm{Pe}\approx (A^3/\mathrm{Ta}^{2/3})^{3/5}(2\tau)^{3/5}$. The largest value of $\mathrm{Ta}$ shown in Figure~\ref{fig:analyticPe_rotation} is just large enough to enter this regime, where $\mathrm{Pe}$ is reduced by rotation at $\tau<1$.  However, this regime is not relevant for the parameter values appropriate for white dwarf and neutron star interiors, and so we do not focus on it here.}. The convective velocities are significantly increased, by a factor of $\mathrm{Ro}^{-1/2}$ (since $\mathrm{Pe}\equiv v_cL/\kappa_T$ is unchanged by rotation, and $L/H_P=\mathrm{Ro}^{1/2}$)\footnote{If using the Rossby number defined with the non-rotating value of $v_c$, the factor by which rotation increases the velocity is $\mathrm{Ro}^{-1/3}$ .}.
For $\tau>1$, the last term in equation \eqref{eq:tau_eq_with_omega} dominates, giving
\begin{equation}\label{eq:Pe_rot}
    \mathrm{Pe}\approx {A^{9/5}\over \mathrm{Ta}^{2/5}}(2\tau)^{3/5},\hspace{1cm}(\tau\ \mathrm{large})
\end{equation}
and
\begin{equation}
    \nabla_X \approx {\chi_T\nabla_\mathrm{ad}\over\chi_X} 
    {\mathrm{Ta}^{2/5}\over A^{9/5}} (2\tau)^{2/5}.\hspace{1cm}(\tau\ \mathrm{large})
\end{equation}
Comparing equations \eqref{eq:largebranch} and \eqref{eq:Pe_rot}, we see that the effect of rapid rotation is to reduce $\mathrm{Pe}$ (increase $\nabla_X$) for $\tau>1$, multiplying (dividing) it by a factor $\approx (A^{4/5}/\mathrm{Ta}^{2/5})(2\tau)^{4/15} \propto \mathrm{Ra_T}^{4/15}/\mathrm{Ta}^{2/5}$ (see Fig.~\ref{fig:analyticPe_rotation}).



\section{Numerical simulations}\label{sec:numerical}

The mixing length theory in the previous section makes a number of approximations and assumptions, in particular in the calculation of thermal losses from convecting fluid elements (eq.~[\ref{eq:nne}]). In this section, we compare against numerical simulations of compositionally-driven convection. We first check that indeed there is an inwards directed heat flux associated with an outwards composition flux. Then, we allow the thermal gradient to come into steady-state and investigate the relation between the composition and thermal gradients and the value of the P\'eclet number that characterizes the flow. 

\subsection{Model and simulation setup}

We conduct simulations for a binary fluid within a 3D spherical shell of depth $\Delta r$. For this first numerical investigation, and to simplify comparison with mixing length theory, we consider a non-rotating system. We express the fluid quantities as the sum of a constant background (denoted by the subscript 0) and a dynamic perturbation to the background (denoted by the prime symbol), e.g., the density $\rho = \rho_{0} + \rho'$. We use the Boussinesq approximation \citep{Spiegel1960}, where density perturbations satisfy $\rho'/\rho_0 \ll 1$, and are related to perturbations in temperature $T'$ and mass fraction of the lighter component $X'$ through $\rho' = -\rho_0 (\beta X' + \alpha T')$, where $\beta$ and $\alpha$ are the coefficients of compositional and thermal contraction/expansion (both assumed positive constants), respectively. Convection is driven by imposing a constant flux of light elements across the domain, such that light elements are injected (removed) at the inner (outer) boundary.

\begin{figure*}
   \centering
    \includegraphics[width=0.92\textwidth]{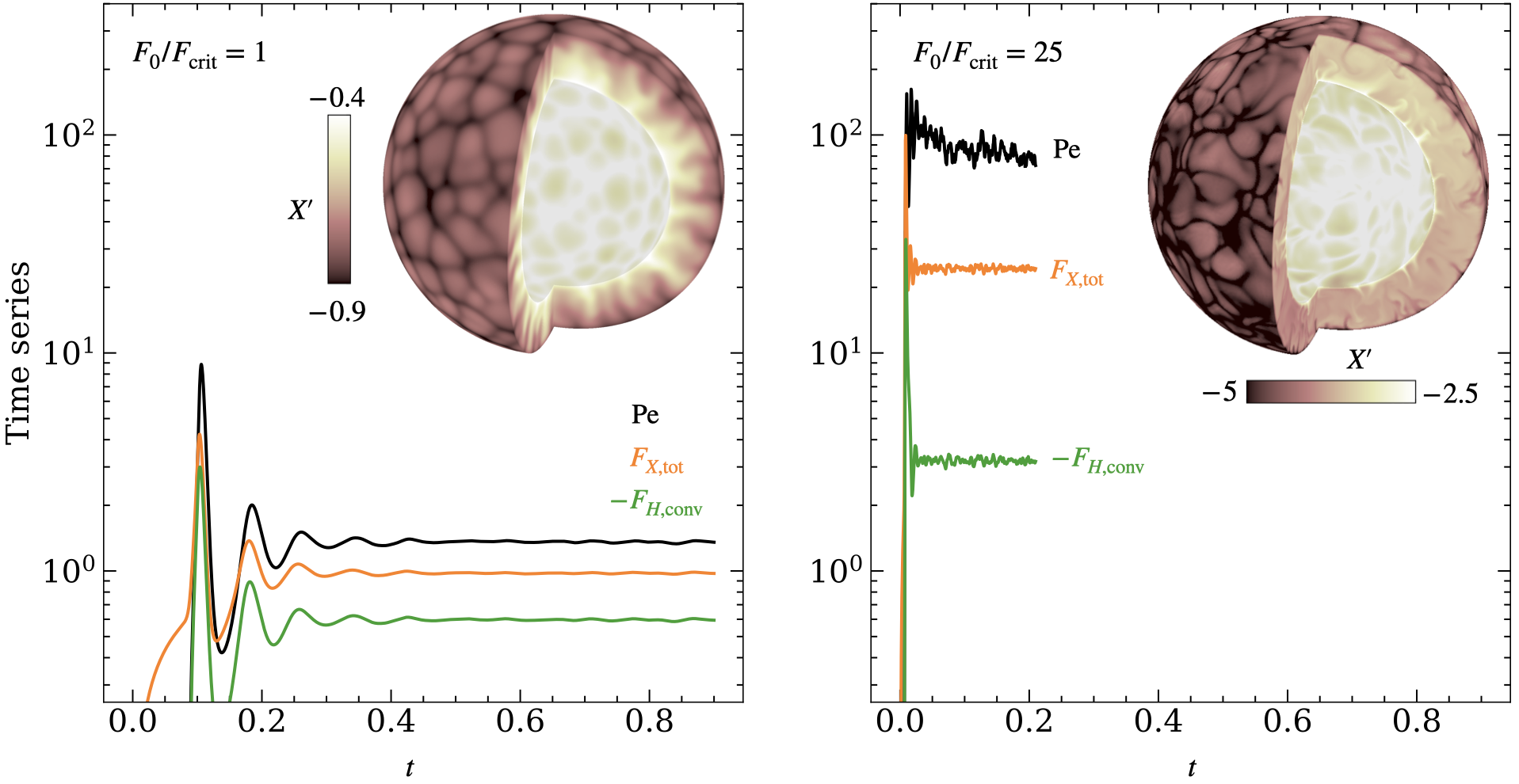}\\
    \includegraphics[width=0.91\textwidth]{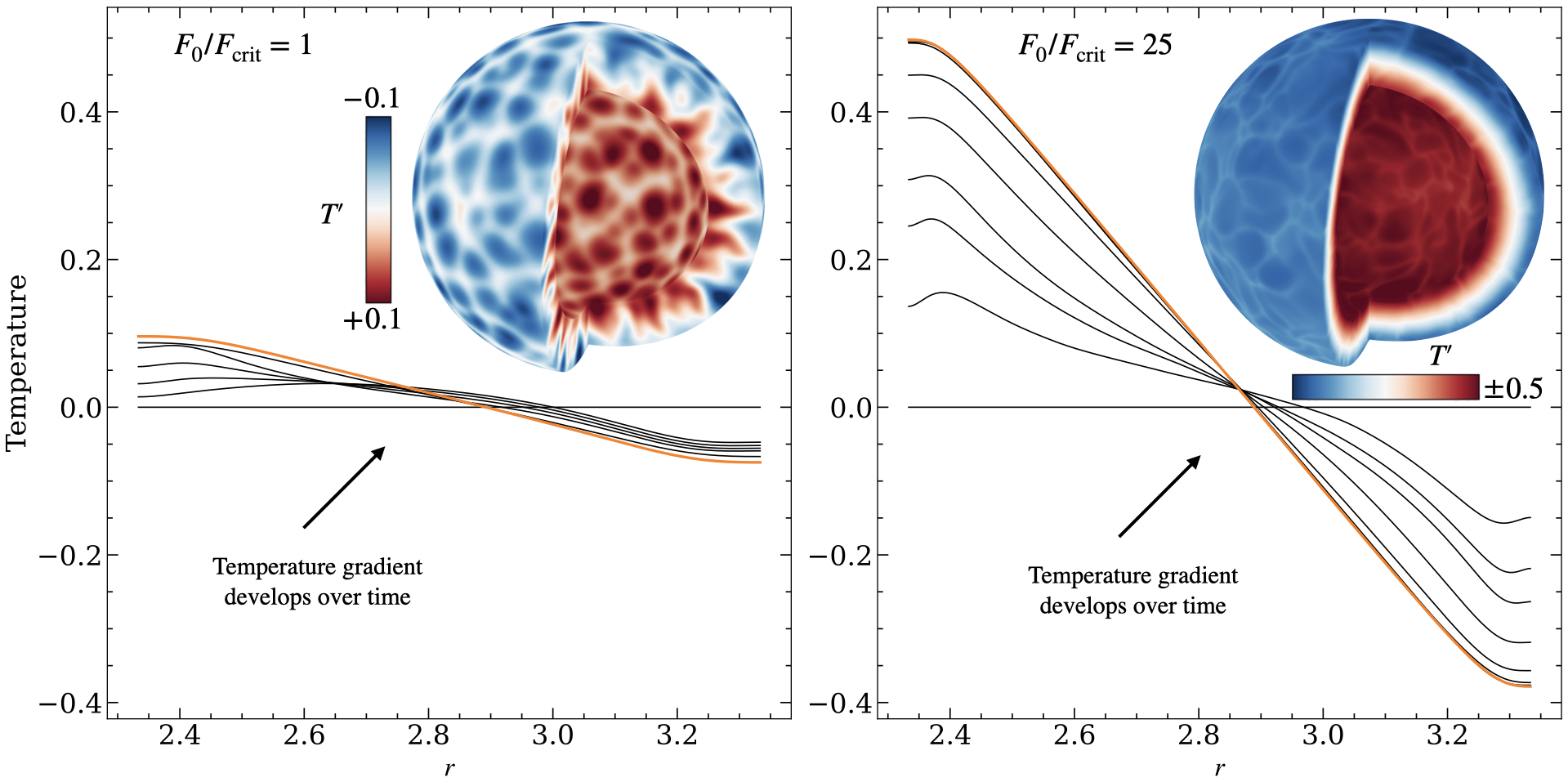}
   \caption{Top panels: Time series of the P\'eclet number, $\mathrm{Pe} = -2\mathrm{Le}^{-1}\langle u_r X' \rangle/\langle \partial_r X' \rangle$ (where the brackets denoting the average over the shell), total composition flux in the radial direction, $F_{X,\mathrm{tot}} = \langle \hat{\bm{r}} \cdot (\bm{u} X' - \nabla X') \rangle$ (where the first term corresponds to the convective flux, and the second term to the diffusion flux), and the magnitude of the convective heat flux in the radial direction, $-F_{H,\mathrm{conv}} = -\langle \hat{\bm{r}} \cdot \bm{u} T' \rangle$. We recall that all quantities are written in dimensionless form. Note that the total composition flux converges to $F_0/F_{\mathrm{crit}}$ once the fluid reaches steady state. Bottom panels: Radial temperature profile at different times. Results are shown for the fiducial cases at low and high $\mathrm{Pe}$, i.e., $F_0/F_{\mathrm{crit}} = 1$ (left panels) and 25 (right panels), respectively. To show the differences in the structure of the flow, we overplot 3D snapshots of the composition and temperature fields, once the simulation reaches steady state.}
  \label{fig:time_series_Pe_flux}
\end{figure*}

We non-dimensionalize the fluid equations using as units of length and time the shell depth, $\Delta r$, and the diffusion time for solute, $t_{\rm diff} = \Delta r^2/\kappa_X$, where $\kappa_X$ is the solute diffusivity. The temperature scale is $[T] = |\partial_r T_{\mathrm{ad}}| \Delta r$, where $\partial_r T_{\mathrm{ad}}$ is the adiabatic temperature gradient. For solute, we use $[X] = (\alpha/\beta) |\partial_r T_{\mathrm{ad}}| \Delta r$. By this choice, a unit of pressure corresponds to $[P] = \rho_0(\kappa_X/\Delta r)^2$, and the corresponding normalizations for the heat and composition flux are $[F_H] = F_{H,\mathrm{ad}}(\kappa_X/\kappa_T)$, and $[F_X] = \rho_0\kappa_X(\alpha/\beta)|\partial_r T_{\mathrm{ad}}| \equiv F_{\mathrm{crit}}$, respectively. Note that $F_{\mathrm{crit}}$ corresponds to the flux of light elements that, if carried by molecular diffusion, would result in a composition gradient that is marginally stable against convection (ie.~$\beta\partial_rX =\alpha|\partial_r T_{\mathrm{ad}}|$).
The dimensionless equations are 
\begin{eqnarray}
&\nabla \cdot \bm{u} = 0\, ,\\
&\dfrac{\partial \bm{u}}{\partial t} +  (\bm{u}\cdot \nabla)\bm{u} = - \nabla P' + \mathrm{Le}^{2}\mathrm{Ra_T}\left(X' + T'\right)\hat{\bm{r}} + \mathrm{Sc} \nabla^2 {\bm{u}}\, ,\\
&\dfrac{\partial X'}{\partial t} +  (\bm{u}\cdot \nabla)X' =  \nabla^2 X'\, ,\\
&\dfrac{\partial T'}{\partial t} +  (\bm{u}\cdot \nabla)T' +  u_r = \mathrm{Le} \nabla^2 T'\, ,
\end{eqnarray}
where we have assumed constant gravity, and $\bm{u} =$ ($u_r$,$u_{\theta}$,$u_{\phi}$) is the velocity field (with $u_r$, $u_{\theta}$ and $u_{\phi}$, the radial, polar, and azimuthal components of the velocity, respectively). In the equations above, there are 3 dimensionless numbers that characterize the evolution of the flow. These are the Rayleigh, Schmidt, and Lewis number, which are defined respectively as
\begin{align}
\mathrm{Ra_T} = \dfrac{ g\alpha \Delta r^4 |\partial_r T_{\mathrm{ad}}|}{\kappa_T^2} \, ,\hspace{0.25cm} \mathrm{Sc} = \dfrac{\nu}{\kappa_X}\, ,\hspace{0.25cm} \mathrm{Le} = \dfrac{\kappa_T}{\kappa_X}\, ,
\end{align}
where $\nu$ is the kinematic viscosity. Note that $\mathrm{Sc} = \mathrm{Pr}\, \mathrm{Le}$, where  $\mathrm{Pr} = \nu/\kappa_T$ is the Prandtl number.

    

\begin{figure}
   \centering
    \includegraphics[width=\columnwidth]{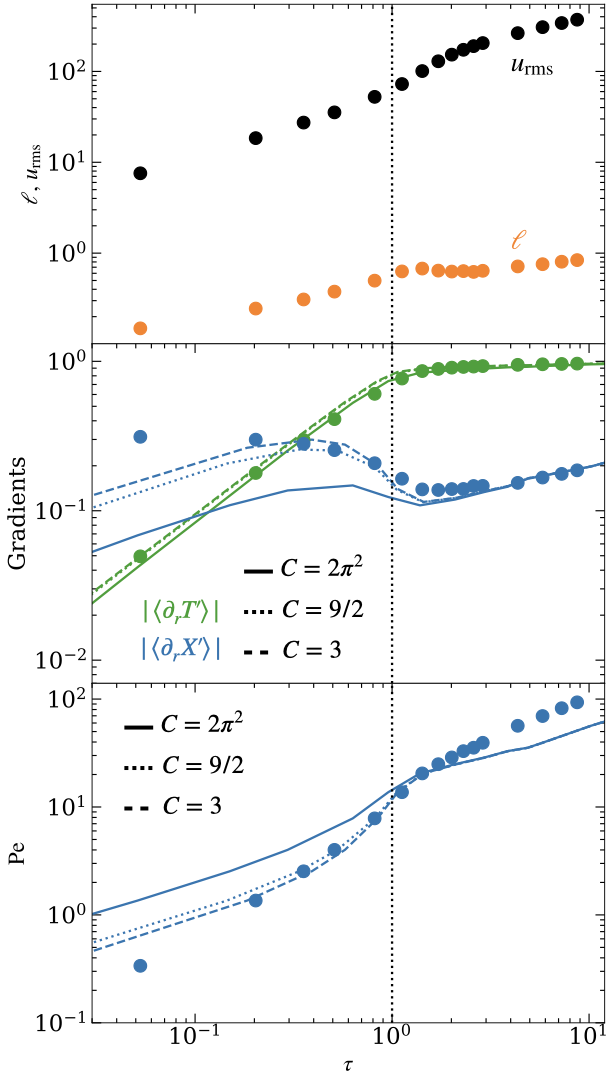}
    \caption{RMS velocity $u_{\mathrm{rms}}$ and mixing length value $\ell$ (top panel), absolute value of the volume-averaged temperature and composition gradient (middle panel) and P\'eclet number (bottom panel) measured from the simulations as a function of the driving parameter $\tau = \mathrm{Le}^{-1}\langle u_r X'\rangle$. Results are shown for simulations using $F_0/F_{\mathrm{crit}} = 0.5$--$30$. Lines on each panel are predictions from the steady-state mixing-length theory solution (Appendix \ref{app:mlt_bouss}) using the measured values of $\ell$ for each case, $\mathrm{Ra_T} = 1.5\times 10^{5}$, and different values of $C$. }
    \label{fig:data_vs_analytic}
\end{figure}

We set the inner and outer radius of the shell to $r_{\mathrm{i}} = 7/3$, and $r_{\mathrm{o}} = 10/3$, respectively. Note that for this choice, the shell depth is $\Delta r = r_{\mathrm{o}} - r_{\mathrm{i}} = 1$, and the aspect ratio is $r_{\mathrm{i}}/r_{\mathrm{o}} = 0.7$. For the dimensionless numbers above, we use $\mathrm{Ra_T} = 1.5\times 10^5$, $\mathrm{Le} = 3.3$, $\mathrm{Pr} = 0.5$, and $\mathrm{Sc} = 1.6$. We selected Pr and Le slightly below and above unity, respectively, and a moderate value of $\mathrm{Ra_T}$, to be in the astrophysical regime \citep[$\mathrm{Pr}\ll 1$, $1 \ll \mathrm{Le} \ll \mathrm{Ra_T}$, e.g.,][]{2019ApJ...876...10S,2021PhRvF...6c0501G} and remain in a numerically-tractable region of the parameter space. The strength of the convective flow is controlled by changing the flux of light elements at the boundaries. The boundary conditions are zero gradient for temperature, and impenetrable and stress-free for velocity ($u_r = \partial_r(u_\theta/r) =  \partial_r(u_\phi/r)=0$). 

We specify the composition flux by setting the value of the composition gradient at each boundary. 
In our dimensionless variables, this is $\partial_r X'\vert_{r=r_{\mathrm{i}},r_{\mathrm{o}}} = -F_0/F_{\mathrm{crit}}$,  where the desired composition flux $F_0$ is normalized by $F_{\mathrm{crit}}$. We consider values of $F_0/F_{\mathrm{crit}}$ between 0.5 and 30.

We solve the governing equations and boundary conditions presented above using the pseudo-spectral solver Dedalus \citep{2020PhRvR...2b3068B,VASIL2019100013,LECOANET2019100012}. The variables are represented in spherical harmonics for the angular directions and Chebyshev polynomials for the radial direction. All the simulations have $L_{\mathrm{max}} = N_{\mathrm{max}} = 255$, where $L_{\mathrm{max}}$ is the maximum spherical harmonic degree, and $N_{\mathrm{max}}$ is the maximal degree of the Chebyshev polynomials used in the radial expansion. Therefore, the number of radial, latitudinal, and longitudinal points are $(N_r,N_\theta,N_\phi) = (256,256,512)$, respectively. For time-stepping, we use a second order semi-implicit BDF scheme \citep[SBDF2,][]{wang_ruuth_2008}, where the linear and nonlinear terms are treated implicitly and explicitly, respectively. We use a CFL safety factor of 0.35 and dealias factor of $3/2$. To start our simulations, we add random noise perturbations to the background composition.  In what follows, all numerical results are presented in dimensionless form.


\subsection{Qualitative description of the flow}

We first present results for the runs using $F_0/F_{\mathrm{crit}} = 1$ and  $F_0/F_{\mathrm{crit}} = 25$ as fiducial cases for low and high P\'eclet number, respectively. We compute the P\'eclet number using the ratio between the convective composition flux and the composition gradient (Eq.~\ref{eq:Pe_bous}), which written in terms of the dimensionless fluxes is $\mathrm{Pe}= -2 \mathrm{Le}^{-1} u_r X'/\partial_r X'$. We find that the behavior is qualitatively similar for all values of $F_0/F_{\mathrm{crit}}$: once the fluxes at the boundaries are turned on, an excess (deficit) of light elements develops at the inner (outer) boundary of the shell. Eventually, the fluid becomes compositionally-buoyant and suddenly overturns, driving convection. 

All the simulations reach a statistically stationary state where the volume-averaged quantities (denoted by brackets $\langle \rangle$) fluctuate around a constant value (see top panels in Fig.~\ref{fig:time_series_Pe_flux}). When computing volume averages, we exclude regions near the diffusive boundary layers and confine our measurements to the convective region. The time to reach steady-state depends on the value of $F_0/F_{\mathrm{crit}}$. For the fiducial cases here, at low $\mathrm{Pe}$ the steady state is achieved at $t\approx 0.5$, whereas at high $\mathrm{Pe}$ it is achieved at $t\approx 0.05$, an order of magnitude difference. We also see differences in the flow structure between the two cases. This can be seen in the 3D snapshots of the composition field in the top panels of Fig.~\ref{fig:time_series_Pe_flux}. We find that the structure of the flow is more diffusive (turbulent) at low (high) $\mathrm{Pe}$.

Our simulations confirm the expected inwards convective heat flux.
We find that for a given composition flux, there is an oppositely directed heat flux that is larger when the composition flux that drives convection is larger (see the green curves in Fig.~\ref{fig:time_series_Pe_flux}). Further, as heat is transported inward, a temperature gradient develops over time until the associated flux carried by diffusion balances the convective heat flux (see bottom panels in Fig.~\ref{fig:time_series_Pe_flux}). This cancellation means that once the simulation reaches steady-state, the total heat flux across the fluid is zero, as expected from our choice of zero flux boundary conditions.

\subsection{Gradients and P\'eclet number in the convective region}
 
As discussed in \S\ref{sec:P\'eclet}, the properties of the flow in the convection zone are expected to change as a function of the driving parameter $\tau$. In particular, mixing-length theory predicts a transition when $\tau = 1$.  To check whether the simulations support this transition, we use the shell-averaged convective velocities and radial fluxes as a function of time, and then for each quantity we take the time-average value over an interval for which the system is statistically stationary. We evaluate $\tau$ using the convective composition flux in equation \eqref{eq:tauBoussinesq}, which written in terms of the dimensionless fluxes gives $\tau = \mathrm{Le}^{-1}(u_r X')$.


Figure \ref{fig:data_vs_analytic} shows the numerical results. We show the rms velocity $u_{\mathrm{rms}}$, measured mixing length $\ell = -2\langle u_r X'\rangle/\langle \partial_r X' \rangle \langle u_{\mathrm{rms}}\rangle$, gradients, and  P\'eclet number as a function of $\tau$ in the top, middle, and bottom panels, respectively. We find that $u_{\mathrm{rms}}$ increases monotonically with $\tau$, and $\ell$ varies between $\approx 0.1$--$0.6$ (becoming approximately constant for $\tau>1$). 
The solid curves in the middle and bottom panels of Figure \ref{fig:data_vs_analytic} are the mixing length theory predictions (which we rewrite for Boussinesq convection in Appendix \ref{app:mlt_bouss}). We use our measured values of $\ell$ and show results for three different values of $C$ reported in the literature (see discussion in \S\ref{sec:leakage}). We find that the data supports the predicted transition at $\tau = 1$, and the general shape of the curves match well. The transition is smoother and shows less of a jump than the example shown in Figure \ref{fig:analyticPe} because of the lower value of Rayleigh number in our simulations. The measured temperature gradient agrees well with the prediction, showing that the magnitude of the convective heat flux is also as predicted. We also see the expected inflection in the dependence of the composition gradient with $\tau$. 

There are some differences between the measured values and the predictions. We find a better agreement for the gradients as a function of $\tau$, than for the P\'eclet number as a function of $\tau$. The values of $\mathrm{Pe}$ are larger than the predicted values for large $\tau$. Fitting separately a power-law to the data gives $\mathrm{Pe} \propto \tau$  for $\tau < 1$ (compared to the analytic prediction $\mathrm{Pe}\propto \tau^{1/2}$), and $\mathrm{Pe}\propto \tau^{0.8}$ for $\tau > 1$ (compared to the analytic prediction $\mathrm{Pe}\propto \tau^{1/3}$). We find that the composition gradient approaches $\partial_r X^\prime\approx 1/\mathrm{Le}\approx 0.3$ at small $\tau$, which is consistent with the expected threshold for double-diffusive instabilities (eg.~\citealt{Traxler2011ApJ}), whereas the analytic model assumes that $\mathrm{Le}$ is large enough that the threshold can be neglected.
We were not able to find values of $C$ that fit all the data points, but smaller values of $C$ are preferred when fitting both $\nabla_X$ (middle panel) and $\mathrm{Pe}$ (bottom panel). Nonetheless, the overall general agreement is encouraging especially given the approximate nature of mixing length theory (particularly the approximations made in deriving eq.~[\ref{eq:Pe}] for the thermal leakage during convection).

\section{Discussion}\label{sec:discussion}

\subsection{Summary of our results}

We have used both mixing length theory and numerical simulations to investigate the heat transport in compositionally-driven convection.
Our results show that there are two different convection regimes, 
depending on the value of the parameter $\tau$ defined in equation \eqref{eq:tau}. When thermal diffusion is very efficient, $\tau\ll 1$, the convective motions have a small P\'eclet number and only a small composition gradient is needed in the convection zone to overcome the reduced thermal buoyancy ($\nabla_X\approx \nabla_{X,\mathrm{crit}}$; eq.~[\ref{eq:DXcrit}]). A small temperature gradient $\nabla\approx \tau\nabla_\mathrm{ad}$ develops in the convection zone to balance the inwards transport of heat due to convection.
When thermal diffusion is inefficient, $\tau\gg 1$, the behavior is very different. The temperature gradient steepens to approach the adiabatic gradient, $\nabla\rightarrow\nabla_\mathrm{ad}$, and the convective heat flux becomes balanced by the outwards conduction along the adiabat\footnote{This regime in which inwards heat transport by convection almost balances the outwards conductive flux along the adiabat has been discussed for the Earth's core, eg.~\cite{Loper1978} and \cite{Labrosse1997}.}. Depending on the size of the composition flux driving convection, the composition gradient in the convection zone can significantly exceed the critical gradient, $\nabla_X>\nabla_{X,\mathrm{crit}}$. 
There is rapid change from one regime to another as $\tau$ crosses unity. 
In both cases, the effect of rapid rotation is to increase the convective velocity and reduce the composition gradient, with only a minor effect on the heat flux or temperature gradient unless the rotation is extremely strong. This behavior is the opposite to what is observed in thermal convection, where rotation reduces the convective velocity \citep{2014ApJ...791...13B,Aurnou2020}.

We find that the ratio of heat flux to composition flux is independent of P\'eclet number at low Pe (eq.~[\ref{eq:FHFXPe}]). 
Rising fluid elements lose heat due to thermal diffusion, reducing the effectiveness of heat transport, but a smaller composition gradient is needed to overcome the thermal buoyancy, reducing the composition transport by the same factor. Our numerical results give support to this scaling. After an initial build up of composition at the boundaries, convection starts and evolves to a state in which, at small P\'eclet number, the gradients in the convection zone take on values that would be stable to the (adiabatic) Ledoux criterion, indicating that thermal diffusion significantly reduces the stratification. 
This can be seen by the fact that $|\partial_r X'|< 1-|\partial_r T'|$ for small $\tau$ in the left panel of Figure~\ref{fig:data_vs_analytic}. This ordering of gradients (Ledoux stable with an unstable composition gradient and stable thermal gradient) corresponds to the regime of fingering or thermohaline convection driven by double-diffusive instabilities (eg.~\citealt{Garaud2021}). Often investigated as the outcome of unstable imposed gradients, in our case the convection is maintained by the continuous injection of elements at the lower boundary, and the gradients develop as a result of the convection.

\subsection{Implications for accreting neutron stars}\label{sec:implications_NS}

The lack of dependence of $F_H/F_X$ on $\mathrm{Pe}$ when Pe $\ll 1$ means that the calculations of \cite{MC11,MC14,MC15} for accreting neutron star oceans used a correct expression for the heat flux even though they assumed adiabatic motion at low Pe. However, the composition gradient is overestimated and convective velocity underestimated in those calculations. For example, whereas the composition gradient that is marginally stable to the Ledoux criterion is given by $\nabla_X/(\nabla_\mathrm{ad}\chi_T/\chi_X)=1$, Figure \ref{fig:analyticPe} for example shows that $\nabla_X/(\nabla_\mathrm{ad}\chi_T/\chi_X)$ ranges from $\approx 10^{-2}$--$0.3$ for $\tau$ in the range $10^{-3}$--$1$, and can be much smaller for $\tau>1$.

The case of accreting neutron stars is interesting because $\tau$ spans a range of values from small to large, covering both convective regimes.
The factor $\chi_X/\chi_T\nabla_{\rm ad}$ is $\sim 30$--$100$ under the degenerate ocean conditions and depends only on the composition at the crystallization depth (see Appendix \ref{app:micro}), so that $\tau\sim (30$--$100) (t_\mathrm{therm}/t_X)$. For cooling following an accretion outburst, the crystallization timescale is comparable to the cooling time, $t_X\sim t_\mathrm{therm}$, so $\tau\sim 30$--$100$. This is consistent with the rapid steepening of the temperature profile seen by \cite{MC14,MC15}. For steady accretion, new crust forms on the accretion timescale, which is $\sim 30\ \mathrm{yr}$ for typical parameters (taking an ocean depth $\approx 10^{13}\ \mathrm{g\ cm^{-2}}$ and accretion rate $10^4\ \mathrm{g\ cm^{-2}\ s^{-1}}$), whereas the thermal timescale is a few days at these depths \citep{Bildsten1995}. Therefore $t_\mathrm{therm}/t_X\sim 3\times 10^{-4}$, giving $\tau\sim 10^{-2}$ for steady accretion.

Even though the neutron star ocean takes years to accrete, it mixes much more rapidly when chemical separation is happening. For a non-rotating star, Figure \ref{fig:analyticPe} gives $\nabla_X/(\nabla_\mathrm{ad}\chi_T/\chi_X)\approx 0.1$ for $\tau\sim 10^{-2}$, implying that the convective velocity is $\approx 10$ times larger than under the adiabatic assumption. With $\mathrm{Pe}\approx 0.3$, the convective turnover timescale $t_\mathrm{conv}=H_P/v_c=t_\mathrm{therm}/\mathrm{Pe}$ at the base of the ocean is a few thermal times ($\sim 10$ days). Rapid rotation reduces this dramatically. Using equation \eqref{eq:Ro} for the Rossby number, the convective turnover time in the rapidly-rotating limit can be written 
\begin{equation}\label{eq:tconv_rot}
    t_\mathrm{conv, rot} = \left({t_\mathrm{therm}\over \mathrm{Pe}}\right)^{2/3}\left(2\Omega\right)^{-1/3}.
\end{equation}
With a rotation period of a few milliseconds, the convective turnover time is $\approx 10\ \mathrm{min}$ (for a scale height $\approx 3000\ \mathrm{cm}$ this corresponds to a convective velocity $v_c\approx 5\ \mathrm{cm\ s^{-1}}$). For cooling neutron stars with $\tau>1$, the convective velocities are even larger. We can evaluate the Rayleigh number with the help of  \cite{Bildsten1995}. From their Eqs. (3.7 and 3.9),  $\nabla_{\mathrm{ad}} \chi_T/\chi_\rho \approx (3/2Z)(k_BT/E_F)$, where $E_F$ is the Fermi energy, and $Z$ is the average atomic number. This approximation assumes an isothermal neutron star ocean and neglect the electron contribution to the entropy gradient. Then, we find $\mathrm{Ra_T}\approx t_\mathrm{therm}^2 (g/H_P)(3/2Z)(k_BT/E_F)\approx 10^{18}$. For non-rotating convection with $\tau>1$, equation \eqref{eq:largebranch} gives $\mathrm{Pe}\approx (\mathrm{Ra_T}\tau)^{1/3}\approx 10^6 \tau^{1/3}$. The convective turnover time is therefore $\approx 0.3\ \mathrm{s}$ (velocity $\approx 0.1\ \mathrm{km/s}$). For $\tau>1$, rapid rotation decreases the convective velocity. With $\mathrm{Ta}=(2\Omega t_\mathrm{therm})^2 \approx 4\times 10^{17}$, equation \eqref{eq:Pe_rot} gives $\mathrm{Pe}\approx 6000\ \tau^{3/5}$, or a turnover time $\approx 1\ \mathrm{min}$ and velocity $\approx 60\ \mathrm{cm\ s^{-1}}$.

Further calculations of the evolution of accreting neutron star oceans would be interesting taking into account our revised estimates of the composition gradients and convective velocities. Mixing on a rapid timescale should have implications for superbursts. These long thermonuclear flashes are thought to be the result of unstable ignition of carbon in the ocean, although significant problems remain in making enough carbon and getting it to ignition temperature \citep{intZand2017}. For example, mixing in the ocean could transport carbon to greater depths where it can burn (stably or unstably).
It would also be interesting to revisit the calculations of \cite{MC14} for neutron stars cooling after accretion outbursts. Recently, \cite{Parikh2020} reported observations of two accreting neutron stars in quiescence that showed a late time ($\approx 2000$ days after outburst) decrease in temperature, followed by a temperature increase. They pointed out that this behaviour is similar to the models of \cite{MC14} that include compositionally-driven convection. Further investigations are needed to compare against the observations for these two sources and explore the constraints on ocean composition and temperature needed to fit the data.

\subsection{Implications for white dwarf cooling and dynamos}\label{sec:implications_WD}

To investigate the parameters for cystallization-driven convection in white dwarfs, we ran an example $0.6\ M_\odot$ white dwarf model using the MESA stellar evolution code \citep{Paxton2011,Paxton2013,Paxton2015,Paxton2018,Paxton2019,Jermyn2022} (using the default \verb|wd_cool_0.6M| test suite in MESA version 22.11.1)\footnote{The MESA inlists and supplement code used in this work are publicly available at\dataset[doi:10.5281/zenodo.7683203]{https://doi.org/10.5281/zenodo.7683203}}. Note that although the code follows the solid-liquid transition and includes the latent heat, it does not include chemical separation and so the composition profile does not evolve in this calculation. Instead, we estimate the composition flux due to chemical separation by measuring the rate of growth of the solid core $\dot M_c$ and assuming a value $\Delta X_\mathrm{melt}=0.1$ for the carbon enhancement in the liquid phase relative to the solid (approximately the liquid-solid composition difference for the C/O phase diagram; \citealt{Horowitz2010}). The composition flux is then $F_X=\dot{M}_c\Delta X_\mathrm{melt}/4\pi R_c^2$, where $R_c$ is the core radius. 

\begin{figure}
   \centering
    \includegraphics[width=1.0\columnwidth]{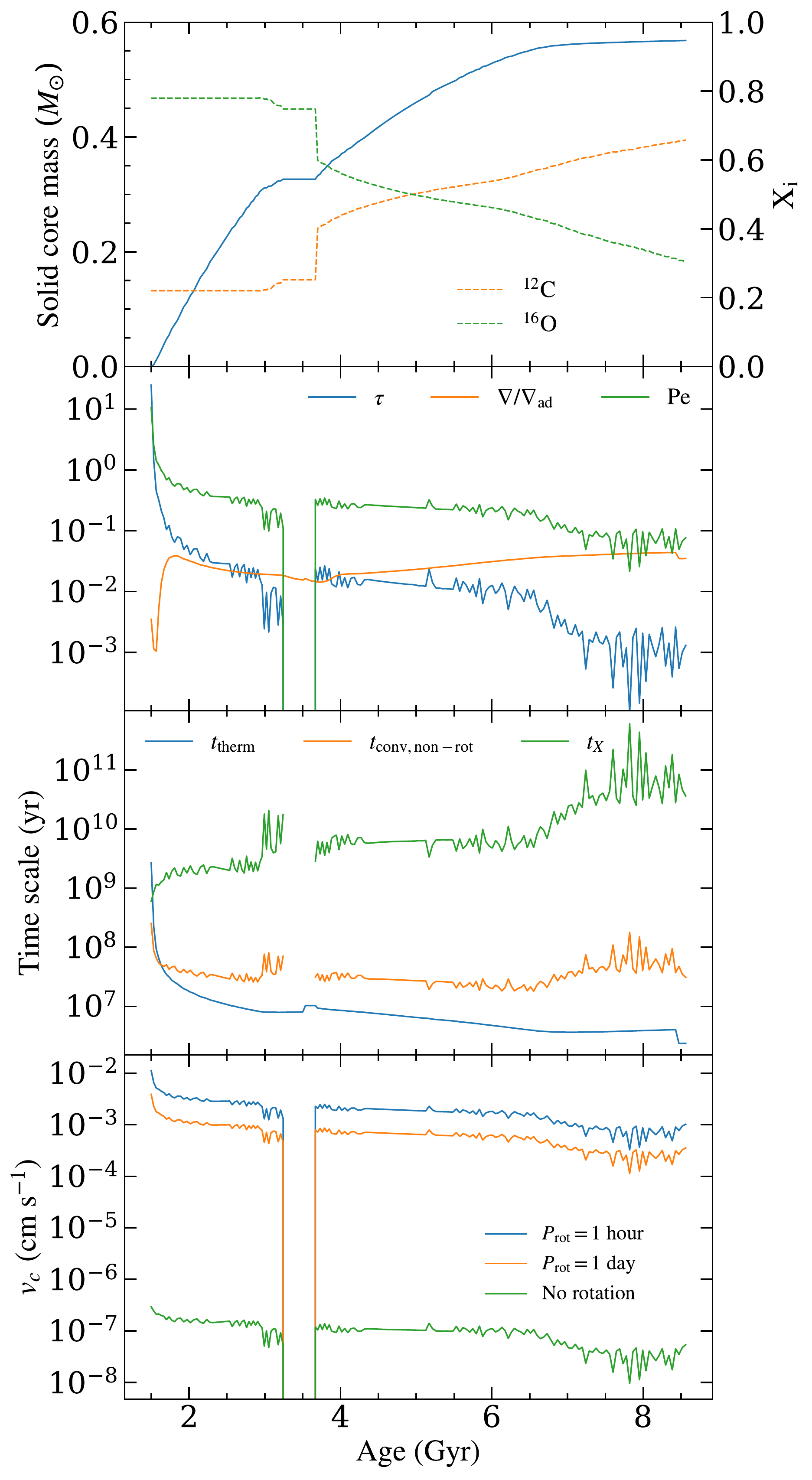}
    \caption{Convection parameters just above the crystallization front as a function of time for a $0.6\ M_\odot$ white dwarf, evolved with the MESA stellar evolution code. Chemical separation and convection are not included in this model, but we use the rate of crystallization of the core to calculate the expected properties of compositionally-driven convection.
    }
    \label{fig:mesawd}
\end{figure}

Figure~\ref{fig:mesawd} shows different parameters associated with the liquid region just above the crystallization front as a function of time. The top panel shows the mass of the solid core and the composition at the freezing point. The white dwarf has an oxygen-rich inner core surrounded by a carbon-rich outer core; growth of the core pauses at $\approx 3\ \mathrm{Gyr}$ when the crystallization front reaches the edge of the inner core; it takes $\approx 0.5\,\mathrm{Gyr}$ of further cooling before the outer core begins to freeze. The values of $t_X$, $t_\mathrm{therm}$, $\tau$ and $\mathrm{Pe}$ and the temperature gradient $\nabla$ are shown in the middle two panels of Figure~\ref{fig:mesawd}. 

As in the neutron star case, $\chi_X/\chi_T\nabla_\mathrm{ad}\sim 10$ is relatively large (see Appendix \ref{app:micro}), but $t_\mathrm{therm}$ in the conductive interior is short enough compared to the evolution time $t_X$ that $\tau$ is small for much of the evolution.
We find $\tau>1$ for a short time at the beginning of crystallization, but it quickly drops and stabilizes at a value of $\tau\approx 0.01$. The corresponding P\'eclet numbers are $\mathrm{Pe}\approx 0.3$, in good agreement with the estimates of \cite{Mochkovitch1983} and \cite{Isern1997}.
The bottom panel of Figure \ref{fig:mesawd} shows the convective velocity. For the non-rotating case, we take $v_c=\kappa_T\mathrm{Pe}/H_P$, and for the rotating case, we use the convective turnover time from equation \eqref{eq:tconv_rot}. This assumes that the mixing length is comparable to the pressure scale height; in our numerical results, we find that the mixing length varies by less than an order of magnitude between small and large $\tau$. At small Pe, our numerical results suggest that the mixing length could be several times smaller than the pressure scale height, which would increase the convective velocity by the same factor

The velocities we obtain are in reasonable agreement with \cite{Mochkovitch1983} who, using a similar formulation of mixing length theory, estimated $v_c\lesssim 10^{-6}\ \mathrm{cm\ s^{-1}}$ for no rotation and $\approx 0.2\ \mathrm{cm\ s^{-1}}$ for a 1 hour rotation period.

Our convective velocities are much smaller than the recent estimates of \cite{Isern2017} and \cite{Ginzburg2022} for crystallization-driven dynamos in white dwarfs. The initial estimates of \cite{Isern2017} considered the acceleration of carbon-rich parcels of fluid released at the phase transition, finding $v_c\approx 30\ \mathrm{km\ s^{-1}}$. \cite{Ginzburg2022} argued that this was an overestimate and instead obtain a velocity $\sim (F_\mathrm{grav}/\rho)^{1/3}$, where $F_\mathrm{grav}$ is the gravitational energy flux associated with the redistribution of elements across the crystallization front. This estimate actually corresponds to the situation where $\tau\gg 1$ and $\nabla_X\gg\nabla_{X,\mathrm{crit}}$. In that case, equation \eqref{eq:vc2} gives $v_c^3\approx gH_P(\chi_X/\chi_\rho)v_c\nabla_X\approx gH_P(\chi_X/\chi_\rho)(F_X/\rho X)\approx F_\mathrm{grav}/\rho$ (eg.~compare equation~\ref{eq:Lgrav}). However, we find that white dwarf interiors are in the $\tau\ll 1$ regime, as previously found by \cite{Mochkovitch1983}, with much lower accelerations and velocities since $\nabla_X\approx \nabla_{X,\mathrm{crit}}$. The change of regime has a huge effect on the velocities: even our rotating convective turnover times are thousands of years, compared to turnover times of months in \cite{Ginzburg2022}. Even with this lower velocity, the magnetic Reynolds number $\mathrm{Rm}$ is likely to be large enough to support a dynamo once rotation is taken into account. 
With the electrical conductivity in the range $\sigma\sim 10^{21}$--$10^{22}\ \mathrm{s^{-1}}$ (eg.~see Fig.~1 of \citealt{Cumming2002}), $\mathrm{Rm}=H_Pv_c/\eta=4\pi\sigma v_c H_P/c^2\sim 10^6$--$10^7$ for $H_P\approx 10^8\ \mathrm{cm}$ and $v_c\approx 10^{-3}\ \mathrm{cm\ s^{-1}}$ appropriate for the rapidly-rotating case (bottom panel of Fig.~\ref{fig:mesawd}). The threshold value of $\mathrm{Rm}$ for a dynamo is uncertain, but with $v_c\propto \Omega^{1/3}$, considering even slower rotation does not reduce $\mathrm{Rm}$ significantly. 

However, another major issue for dynamos is the energy reservoir available to grow the field. In Appendix \ref{app:ener}, we show that the kinetic energy flux is a small fraction of the available gravitational energy. 
The saturated dynamo scaling used by \cite{Isern2017} is $B^2/4\pi\sim \rho v^2$ with $v\sim (F/\rho)^{1/3}$ \citep{Christensen2009}, where they assumed that the energy flux $F$ available to drive the dynamo was the gravitational energy flux $F_\mathrm{grav}$. 
The mechanism by which dynamo saturation occurs and the force balance in the saturated state is still an area of active study \citep{Christensen2006,Schaeffer2017,Orvedahl2021}. However, since magnetic field generation occurs as a result of induction by fluid motions, it seems unlikely that the magnetic energy density could be many orders of magnitude larger than the kinetic energy of the flow. In the context of the Earth's core, \cite{Loper1978} also pointed out that much less kinetic energy is available to drive the dynamo when compositionally-driven convection occurs in a thermally-stable background. To estimate how small this is, we can use equation \eqref{eq:fk_fh_low_tau}. Assuming a solid core mass $\sim 0.1\ M_\odot$ and $H_P\sim 10^8\ {\rm cm}$, we find $\mathrm{Ra_T}\sim 10^{28}$ and $\mathrm{Ta}\sim 10^{24}\, (P_{\rm rot}/{\rm h})^{-2}$, giving 
$F_K/F_H\sim \mathrm{Ta}^{1/3}/\mathrm{Ra_T}\sim 10^{-20}$. Using $B^2/4\pi \sim \rho v^2$ with $v_c\sim 10^{-3}\ {\rm cm\ s^{-1}}$ gives $B\sim 3\ {\rm G}\ \rho_6^{1/2}$, much smaller than needed to explain observed magnetic fields in white dwarfs.

The results in Figure~\ref{fig:mesawd} show that the temperature gradient needed to balance the inwards convective transport of heat ($\approx \tau\nabla_\mathrm{ad}$ for small $\tau$) is larger or comparable in size to the existing temperature gradient in the cooling model for much of the early evolution. This can be seen in the second panel of Figure~\ref{fig:mesawd} where, between $\approx 2$--$6\ \mathrm{Gyr}$, the temperature gradient in the white dwarf normalized to the adiabatic gradient, $\nabla/\nabla_\mathrm{ad}$, is comparable to the value of $\tau$. This is consistent with the significant contribution that chemical separation makes to white dwarf cooling curves. Chemical separation is typically included in white dwarf cooling codes by assuming that the cooling is slow enough that the liquid region is well-mixed \citep{Isern1997,Isern2000, Salaris1997, Montgomery1999}. The energy change due to the changing composition profile is then added to the latent heat, and distributed in a small region around the crystallization front \citep{Althaus2010,Camisassa2019,Bedard2022}. This additional energy will lead to a steepening of the temperature gradient (to conduct the extra heat to the surface), and indeed we estimate in Appendix \ref{app:ener} that the magnitude of the convective heat flux is comparable in magnitude to the overall energy release due to chemical separation. This suggests that the temperature profile including the detailed transport of heat associated with mixing above the crystallization front may not be that different from current models, but further calculations are needed to check this in detail. Of particular interest is the beginning of crystallization, when $\tau>10$ and there is the possibility of significant steepening of the temperature gradient in the central regions of the star.

\subsection{Future work on compositionally-driven convection}

The agreement between our numerical simulations and the mixing-length theory predictions shown in \S\ref{sec:numerical} is encouraging. There are many interesting questions to address with further numerical simulations. The value of Rayleigh number that we used in \S\ref{sec:numerical} gives a relatively smooth transition between the small and large $\tau$ regimes (Fig.~\ref{fig:data_vs_analytic}). Simulations at larger Rayleigh number would be interesting to check the rapid transition predicted at $\tau=1$ for large $\mathrm{Ra_T}$. Our mixing length theory results including thermal diffusion provide a convenient interpolation between the fingering and overturning convection regimes. At low $\mathrm{Pe}$, they agree with earlier analytic prescriptions for thermohaline convection (\citealt{Ulrich1972} and \citealt{Kippenhahn1980} as implemented in the MESA code for example, \citealt{Paxton2013}). However, more recent results are available which provide composition and heat fluxes for fingering convection that are measured directly from numerical simulations \citep{Traxler2011ApJ,Traxler2011JFM,Brown2013}. It would improve the modelling to incorporate these results at low $\mathrm{Pe}$.

Even more important is that our simulations do not include rotation, and also adopt the Bousinessq approximation which limits the vertical scale to be much less than a pressure scale height. Rapid rotation should greatly reduce the lengthscale of convection perpendicular to the rotation vector, and is important to check numerically. Similarly, stratification over many pressure scale heights would be expected to limit the vertical transport. Dynamos in fingering convection are beginning to be addressed with numerical simulations. \cite{Mather2021} simulated dynamos with internal volumetric sources or sinks of both thermal and compositional buoyancy, and did not find dynamo action in the fingering convection regime, although \cite{Guervilly2022} argues that fingering convection could support a dynamo at larger Rayleigh numbers. Numerical simulations of compositionally-driven dynamos with a thermally-stable background are needed for application to white dwarfs. It will also be interesting to investigate other sources of compositional buoyancy, for example the distillation process involving production of light crystals proposed by \cite{Blouin2021} for white dwarfs, or electron captures in neutron star oceans that produce heavy crystals within the liquid layer that then sink \citep{MC14} (an analagous case in planetary dynamos is the iron snow in Ganymede's core; \citealt{Ruckriemen2015}). 
These improvements in numerical modelling are needed to interpret the rich set of observations of both white dwarfs and neutron stars now available.

\begin{acknowledgements}
We thank Simon Blouin whose question about the effect of thermal diffusion on compositionally-driven convection sparked this investigation, Brad Hindman and Nick Featherstone for useful conversations on rotating convection. We also thank Thomas Villeneuve and Charles Wilson for preliminary work on this problem. We thank the referee for a thorough and insightful report that improved the paper. This work was supported by NSERC Discovery Grant RGPIN-2017-04780, and NASA through grants 80NSSC19K0267 and 80NSSC20K0193. J.R.F. acknowledges support from a McGill Space Institute (MSI) Fellowship. A. C., J. R. F. and M. C.-T. are members of the Centre de Recherche en Astrophysique du Québec (CRAQ) and the Institut de recherche sur les exoplanètes (iREx). EHA was supported by a CIERA Postdoctoral Fellowship. This research was enabled in part by support provided by Calcul Québec (calculquebec.ca), and Compute Canada (www.computecanada.ca). Computations were performed on Graham and Béluga.
\end{acknowledgements}

\bibliographystyle{aasjournal}
\bibliography{notes} %

\appendix
\section{Mixing length theory for Boussinesq convection} \label{app:mlt_bouss}

In this Appendix, we give the mixing-length theory results from \S\ref{sec:MLT} in a form appropriate for comparison with our numerical results in \S\ref{sec:numerical}, ie.~in terms of the spatial gradients and using the Boussinesq equation of state. The convective fluxes are
\begin{align}
&F_H \approx \dfrac{1}{2}\rho_0 v_c c_P \ell \left(\partial_r T_{\mathrm{ad}}-\partial_r T\right) \dfrac{\mathrm{Pe}}{C + \mathrm{Pe}}\,,\\ 
&F_X \approx -\dfrac{1}{2}\rho_0 v_c \ell \partial_r X\, , \label{eq:fluxes_comp_bous}
\end{align}
with
\begin{equation}
v_c^2 \approx \dfrac{g\ell^2}{8} \left[\alpha \left(\partial_r T_{\mathrm{ad}}-\partial_r T \right){\mathrm{Pe}\over C + \mathrm{Pe}} - \beta \partial_r X\right]\, .
\end{equation}
The minus sign in the definition of $F_X$ takes into account the fact that decreasing composition with radius, $\partial_r X<0$, leads to an outwards composition flux, $F_X>0$. Note that from Eq.~\eqref{eq:fluxes_comp_bous} we can write the mixing length $\ell$
\begin{equation}
\ell \approx \dfrac{-2F_X}{\rho_0 v_c \partial_r X}\,,
\end{equation}
which can be measured directly from the simulations using $v_c = u_{\mathrm{rms}}$, the rms flow velocity.

From the equations above, the equivalent to equations \eqref{eq:DXcrit}, \eqref{eq:XminusXcrit}, \eqref{eq:nablass} and \eqref{eq:evalPe} are
\begin{align}\label{eq:equivalent_bouss}
&\partial_r X_{\mathrm{crit}} \approx {\alpha\over \beta} (\partial_r T_{\mathrm{ad}} - \partial_r T) {\mathrm{Pe}\over C + \mathrm{Pe}},\\
&\partial_r X - \partial_r X_{\mathrm{crit}} \approx  \dfrac{8}{\mathrm{Ra_T}}\dfrac{\alpha \partial_r T_{\mathrm{ad}}}{\beta}\mathrm{Pe}^2\left(\dfrac{\Delta r}{\ell}\right)^4\,,\\
&\partial_r T \approx \partial_r T_{\mathrm{ad}}\left(\dfrac{\mathrm{Pe}^2}{\mathrm{Pe}^2 + 2\mathrm{Pe} + 2C}\right)\, \\
&\mathrm{Pe} \approx \dfrac{-2 F_X}{\rho_0 \partial_r X \kappa_T}\approx \dfrac{t_{\mathrm{therm}}}{t_X}\left(\dfrac{-2X}{\partial_r X \Delta r}\right)\, , \label{eq:Pe_bous}
\end{align}
where $\mathrm{Ra_T} = \alpha g|\partial_r T_{\mathrm{ad}}| \Delta r^4 / \kappa_T^2$, $t_{\mathrm{therm}} = \Delta r^2/\kappa_T$, and $t_X = \rho_0 X \Delta r /F_X $. Following the same argument as in \S\ref{sec:MLT}, we solve the system of equations above in terms of the driving parameter
\begin{equation}\label{eq:tauBoussinesq}
\tau = \left(\dfrac{t_{\mathrm{therm}}}{t_X}\right)\left(\dfrac{-\beta X}{\alpha \partial_r T_{\mathrm{ad}}\Delta r}\right) 
\end{equation}
The expressions above are used to generate the analytic curves in Figure \ref{fig:data_vs_analytic}.

\section{Microphysics of white dwarf interiors and neutron star oceans} \label{app:micro}

In this Appendix, we estimate the expected size of the ratio $\chi_X/\chi_T\nabla_\mathrm{ad}$ that enters into the parameter $\tau$ (eq.~\ref{eq:tau}). For simplicity, as in the main text we consider a mixture of two species only, although it is straightforward to generalize to additional species if needed. The pressure has a contribution from electrons and ions, $P=P_e+\sum_{i=1}^2 P_i$, where the terms with $i=1$ and $i=2$ are the ion contributions from each species. Under the degenerate conditions in white dwarf and neutron star interiors, the degenerate electrons dominate the pressure, with Fermi momentum $p_F=\hbar(3\pi^2n_e)^{1/3} = xm_ec$ given by $x = 1.01\ (\rho_6Y_e)^{1/3}$ where $\rho_6 = \rho/10^6\ {\rm g\ cm^{-3}}$ and $n_e$ is the electron number density. For non-relativistic electrons ($x\ll 1$), the pressure is $P_e=(2/5) n_eE_F$, with $E_F=p_F^2/2m_e$. Therefore, $P_e\propto (\rho Y_e)^{5/3}$, so that $\chi_\rho=5/3$. For relativistic electrons ($x\gg 1$), $E_F=p_Fc$, $P_e\propto (\rho Y_e)^{4/3}$, giving $\chi_\rho=4/3$. 

 Both the electrons and ions play a role in setting the compositional dependence of the pressure, with the dominant contribution coming from the zero-temperature terms. 
For a two-component mixture, 
\begin{equation}
    Y_e = {XZ_1\over A_1} + {(1-X)Z_2\over A_2},
\end{equation}
where $X=X_1$ is the mass fraction of the lighter species, and $1-X=X_2$ is the mass fraction of the heavier species. 
This gives 
\begin{equation}\label{eq:dPedX}
    {\partial P_e\over \partial X} = {\partial P_e\over\partial Y_e}{\partial Y_e\over\partial X} = {\partial P_e\over\partial Y_e}\left[{Z_1\over A_1}-{Z_2\over A_2}\right],
\end{equation}
where $\partial \ln P_e/\partial \ln Y_e=5/3$ and $4/3$ in the non-relativistic and relativistic limits respectively, and the partial derivatives are taken at constant temperature and density. For ions in the liquid phase, the leading order term in the Helmholtz free-energy at zero-temperature contributed by each species $i$ is $F_i = -C_M\,\Gamma_i N_ik_BT$, for $N_i$ ions in a volume $V$, where $\Gamma_i = Z_i^{5/3}\Gamma_e$ and $\Gamma_e=e^2/a_ek_BT$ with $4\pi a_e^3 n_e/3 =1$ defines the mean electron separation $a_e$, and $C_M\approx 0.90$ is related to the Madelung constant \citep{Dewitt1999, Farouki1993, Potekhin2000, MC10}. We therefore find $F_i\propto V^{-1/3}$ leading to the Coulomb pressure $P_i = (1/3)(F_i/V)$, or
\begin{equation}\label{eq:Pi}
P_i = -{1\over 3} C_M n_i Z_i^{5/3}e^2 \left({4\pi n_e \over 3}\right)^{1/3} = -{1\over 3} C_M e^2 \left({4\pi\over 3}\right)^{1/3}\left({\rho\over m_p}\right)^{4/3} Z_i^{5/3} \left({X_i\over A_i}\right) Y_e^{1/3}.
\end{equation}
Therefore,
Then, assuming linear mixing so the pressure contributions from each species add,
\begin{equation}\label{eq:dP12dX}
{\partial (P_1+P_2)\over\partial X} = {P_1\over X} -{P_2\over 1-X} + {1\over 3}{P_1+P_2\over Y_e}{\partial Y_e\over \partial X}  =  {P_1\over X} -{P_2\over 1-X} + {1\over 3}{P_1+P_2\over Y_e}\left[{Z_1\over A_1}-{Z_2\over A_2}\right].
\end{equation}
Now adding the ion and electron contributions (eqs.~[\eqref{eq:dPedX}] and [\ref{eq:dP12dX}]) gives
\begin{equation}
    {\partial P\over\partial X} = {P_1\over X} -{P_2\over 1-X} + \left[{1\over 3}{P_1+P_2\over Y_e}+{\partial P_e\over\partial Y_e}\right]\left[{Z_1\over A_1}-{Z_2\over A_2}\right].
\end{equation}
Noting that $|P_1+P_2|\ll P_e$, we can drop the $P_1+P_2$ term relative to the $\partial P_e/\partial Y_e$ term, and take $P\approx P_e$, giving
\begin{equation}\label{eq:anal_chiX}
    \chi_X \approx -{P_1\over P}\left[\left({Z_2\over Z_1}\right)^{5/3}\left({A_1\over A_2}\right)-1\right] + {X\over Y_e}{\partial \ln P_e\over\partial \ln Y_e}\left[{Z_1\over A_1}-{Z_2\over A_2}\right]
\end{equation}
as our final expression for $\chi_X$. A similar expression for the internal energy per gram $E$ was derived by \cite{Isern1997,Isern2000}; as a check, calculating $\chi_X$ as $(X/P)(\rho^2\partial/\partial\rho)(\partial  E/\partial X)$ using their results for $\partial E/\partial X$ gives agreement with equation \eqref{eq:anal_chiX}.

\begin{figure*}
   \centering
    \includegraphics[width=1.0\columnwidth]{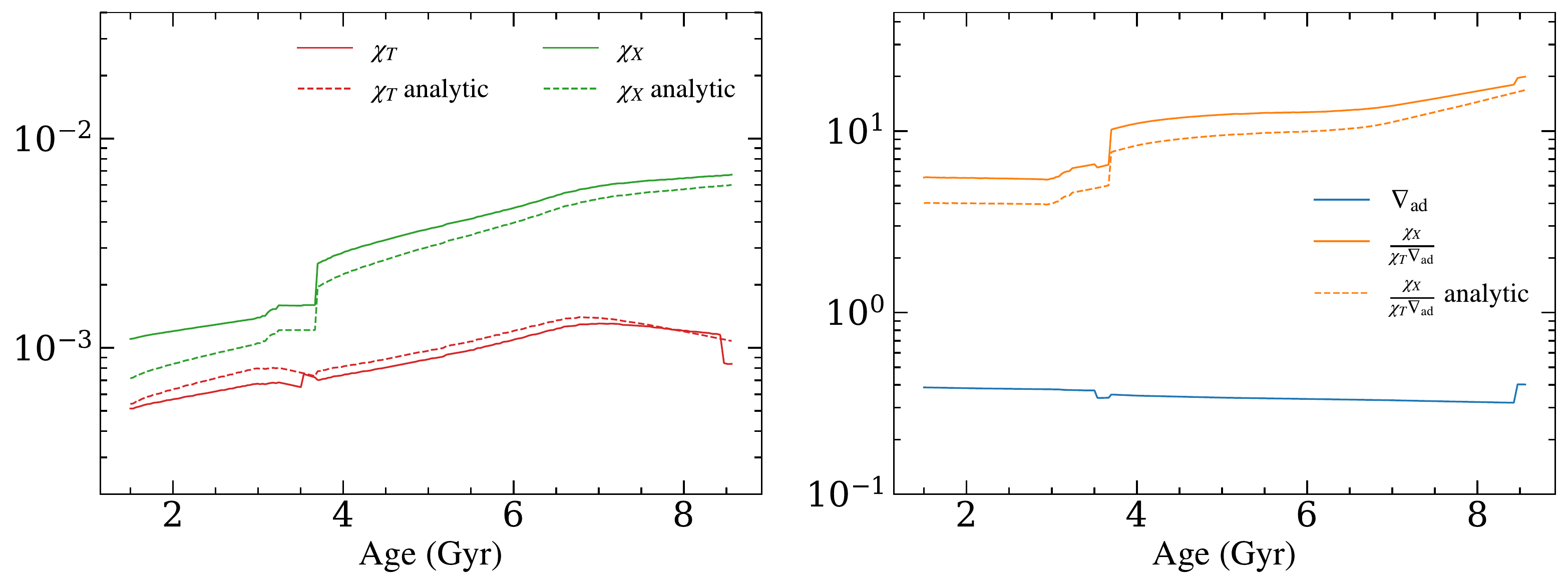}
    \caption{$\chi_X$, $\chi_T$, $\nabla_\mathrm{ad}$, and the ratio $\chi_X/\chi_T\nabla_\mathrm{ad}$ at the crystallization front for the white dwarf models shown in Figure \ref{fig:mesawd}. We compare against the analytic results given by equation \eqref{eq:chi_T_NR}, the first term of equation \eqref{eq:anal_chiX}, and equation \eqref{eq:WDratio}.
    }
    \label{fig:chis}
\end{figure*}

The temperature-dependence of the pressure is dominated by the temperature-dependence of the ion pressure. For degenerate electrons, $\partial\ln P_e/\partial\ln T \sim (k_BT/E_F)^2$, which is much smaller than the ion contribution. The leading temperature-dependent pressure term for the ions is the ideal gas pressure $P_i = n_i k_BT = \rho X_i k_BT / A_i m_p$. However, this receives a significant correction from Coulomb interactions. The Madelung term in the free energy is temperature-independent, but higher order terms do depend on temperature. Writing the first such term in the free energy as a power law $C_T N_ik_BT \Gamma^a$, we find
\begin{equation}
    \chi_T \approx {\rho k_BT\over \mu_i m_p P}\left[1 + {1\over 3}C_T a(1-a)\Gamma^a\right],
\end{equation}
where $\mu_i^{-1}=\sum_i X_i/A_i$. Using the values $C_T=1.865$ and $a=0.323$ from \cite{Dewitt2003} (see equation (2) of \citealt{MC10}) gives $C_T a(1-a)\Gamma^a/3=0.721\ (\Gamma/175)^{0.323}$, where $\Gamma\approx 175$ is the crystallization point for a one-component plasma \citep{Potekhin2000}. Since this Coulomb correction factor depends weakly on $\Gamma$, we take the value at $\Gamma=175$ for simplicity to evaluate $\chi_T$:
\begin{equation}\label{eq:chi_T_NR}
    \chi_T \approx {5\over 2} {k_BT\over E_F} Y_e^{-1} \sum_i {X_i\over A_i}\times 1.721\approx 1.45\times 10^{-3}\ T_6(Y_e\rho_6)^{-2/3}Y_e^{-1}\sum_i {X_i\over A_i}\hspace{1cm}x\ll 1
\end{equation} 
and
\begin{equation}\label{eq:chi_T_R}
    \chi_T \approx   4 {k_BT\over E_F} Y_e^{-1} \sum_i {X_i\over A_i}\times 1.721\approx 1.16\times 10^{-2}\ T_8\rho_9^{-1/3}Y_e^{-4/3}\sum_i {X_i\over A_i}\hspace{1cm}x\gg 1.
\end{equation}
The adiabatic gradient is given by $\nabla_\mathrm{ad}=\chi_TP/\rho c_VT\Gamma_1$. To obtain the correction to the ideal gas heat capacity for the ions we must again go to the next order terms in the Helmholtz free energy since the Madelung term is temperature-independent. Using the same power law free energy as above, we find $c_V\approx (3/2)(k_B/\mu_i m_p)(1+2C_Ta(1-a)\Gamma^a/3)$. Evaluating $\nabla_\mathrm{ad}$, we see that it depends very weakly on density or $\Gamma$. For the same values of $a$ and $C_T$ as above, we find $c_V\approx 3.7\ k_B/\mu_i m_p$ and $\nabla_\mathrm{ad}\approx 0.28$ for non-relativistic electrons ($\Gamma_1=5/3$) and $0.35$ for relativistic electrons ($\Gamma_1=4/3$) at $\Gamma=175$. Other terms in the free energy contribute and so this slightly underestimates $\nabla_\mathrm{ad}$ (see Figure \ref{fig:chis}). For simplicity below we take $\nabla_\mathrm{ad}$ to be constant equal to $1/3$.

We can now estimate the ratio $\chi_X/\chi_T\nabla_{\rm ad}$. In neutron star oceans, the second term in equation \eqref{eq:anal_chiX} dominates, since $P_i\ll P$ and we typically have species with different ratios $Z/A$. 
With $Z_2/A_2<Z_1/A_1$ (since species 1 is the lighter species), this term is positive. For example, for the neutron star ocean with a mixture of O ($Z=8$, $A=16$) and Se ($Z=34$, $A=79$) considered by \cite{MC11}, $\Delta (Z/A)= 0.070$, and the second term gives $\chi_X\approx 0.2 X$. Taking $\nabla_\mathrm{ad}\approx 1/3$ and using equation \eqref{eq:chi_T_R}, we find 
\begin{equation}\label{eq:NSratio}
    {\chi_X\over \chi_T\nabla_\mathrm{ad}}\approx 157\ \left({X\over 0.1}\right) \left({\mu_i\over 79}\right) \left({\Delta(Z/A)\over 0.07}\right){\rho_9^{1/3}\over T_8}
    \approx 39\ \left({X\over 0.1}\right) \left({\mu_i\over 79}\right)\left({\Delta(Z/A)\over 0.07}\right) \left({\langle Z^{5/3}\rangle\over 34^{5/3}}\right)^{-1}\left({\Gamma\over 175}\right),\hspace{1cm}\mathrm{neutron\ star}
\end{equation}
where $\mu_i^{-1} = \sum_i (X_i/A_i)$ and $\Gamma$ is the Coulomb coupling parameter with $\langle Z^{5/3}\rangle$ averaged by number (see \citealt{MC15} eq.~[2]). Note that at fixed $\Gamma$, $\chi_X/\chi_T\nabla_\mathrm{ad}$ depends only on composition and is independent of temperature and density.

In white dwarf interiors, however, the electron term in equation \eqref{eq:anal_chiX} is small or vanishing (as pointed out by \citealt{Isern1997,Isern2000}). For a mixture of C and O for example, $Y_e$, and therefore the electron pressure, is independent of the C/O ratio, since both species have $A=2Z$. In that case, $\chi_X$ is set by the ion term. Using $P_e$ for non-relativistic electrons and $P_i$ from equation \eqref{eq:Pi}, we find $P_i/P_e\approx -0.0057\  (Y_e \rho_6)^{-1/3}  ({Z_i/Y_eA_i}) X_i   Z_i^{2/3}$. For a mixture of C/O, $P_i/P_e\approx 0.019 X(Y_e\rho_6)^{-1/3}$ and the factor $(Z_2/Z_1)^{5/3}(A_1/A_2)-1\approx 0.21$, giving $\chi_X\approx 4.0\times 10^{-3}\ X(Y_e\rho_6)^{-1/3}$, approximately two orders of magnitude smaller than in the neutron star case. Again taking $\nabla_\mathrm{ad}\approx 1/3$, and using equation \eqref{eq:chi_T_NR}, we find
\begin{equation}\label{eq:WDratio}
    {\chi_X\over \chi_T\nabla_\mathrm{ad}}\approx 23 \,\left({X\over 0.5}\right)\left({\mu_i\over 14}\right){\rho_6^{1/3}\over T_6} 
    \approx 8.0\,\left({X\over 0.5}\right)\left({\mu_i\over 14}\right)\left({\langle Z^{5/3}\rangle\over 7^{5/3}}\right)^{-1}\left({\Gamma\over 175}\right).\hspace{1cm}\mathrm{white\ dwarf}
\end{equation}
We see that $\chi_X/\chi_T\nabla_\mathrm{ad}$ is about an order of magnitude smaller than in the neutron star ocean case at the same value of $X$, but still larger than one. We apply these values of $\chi_X/\chi_T\nabla_\mathrm{ad}$ in our estimates in \S \ref{sec:discussion}. 

For the MESA simulation results shown in \S\ref{sec:discussion}, we take $\chi_T$ directly from the code, and compute $\chi_X$ by perturbing $X$ and calling the equation-of-state directly to compute $\partial P/\partial X$. The analytic formulae above agree well with the numerical results, as can be seen in Figure \ref{fig:chis}.

\section{Energetics of chemical separation in white dwarfs} \label{app:ener}

By considering the change of internal energy $E$ with composition across the white dwarf, \cite{Isern1997,Isern2000} found the extra luminosity generated by the redistribution of elements in the convection zone is given by
\begin{equation}
    L_\mathrm{chem} = \dot M_c \Delta X_\mathrm{melt} \left[\left.{\partial E\over \partial X}\right|_c - \left\langle{\partial E\over\partial X}\right\rangle\right] = \dot M_c \Delta X_\mathrm{melt} \, \alpha \left.{\partial E\over \partial X}\right|_c,
\end{equation}
where $\dot{M}_c$ is the growth rate of the mass of the solid core. The partial derivative $\partial E/\partial X$ is taken at constant $T$ and $\rho$; for clarity we do not indicate this explicity. The first term in the square brackets is evaluated at the crystallization boundary, the second is an average over the liquid region, and we introduce the same averaging parameter $\alpha\lesssim 1$ as \cite{Isern1997}. Note that as elsewhere in this paper $X$ is the mass fraction of the light element, and we define $\Delta X_\mathrm{melt}=X_l-X_s>0$ as the difference in the light element mass fraction between liquid and solid phases. Now using equation (8) of \cite{Isern2000} for $\partial E/\partial X$ and the first term in equation \eqref{eq:anal_chiX} for $\chi_X$, we find $\partial E/\partial X = 3gH_P \chi_X/X$, and therefore
\begin{equation}
    L_\mathrm{chem} \approx \dot M_c gH_P\,\Delta X_\mathrm{melt} \,\left(\dfrac{3\alpha \chi_X}{X}\right)
\end{equation}
(see \citealt{Isern1997} for a similar argument). \cite{Ginzburg2022} estimated the rate of gravitational energy release (see their eq.~[6]) as
\begin{equation}
L_\mathrm{grav} \approx  \dot{M}_c gH_P\, {\Delta\rho\over\rho},
\end{equation}
where $\Delta\rho=\rho_s-\rho_l>0$ is the density contrast between solid and liquid phases at the crystallization front. 
Now writing $\Delta\rho/\rho=-(\chi_X/\chi_\rho)(-\Delta X_\mathrm{melt}/X)$ gives
\begin{equation}\label{eq:Lgrav}
L_\mathrm{grav} \approx  \dot{M}_c gH_P\,\Delta X_\mathrm{melt} \,{\chi_X\over X\chi_\rho}\approx \dot{M}_c gH_P\,\Delta X_\mathrm{melt} \,\left({3\chi_X\over 5X}\right),
\end{equation}
which is approximately equal to $L_\mathrm{chem}$ (depending on the value of $\alpha$). 

We can compare $L_\mathrm{chem}$ with the convective heat flux associated with the flux of light elements using equation \eqref{eq:FHFXPe}. Writing $4\pi R_c^2 F_X = \dot M_c\Delta X_\mathrm{melt}$ and assuming $\tau<1$ so that $\nabla_X\approx\nabla_{X,\mathrm{crit}}$, gives
\begin{equation}
 L_H = 4\pi R_c^2 F_H = 4\pi R_c^2 F_X {c_PT\over X} {\chi_X\over\chi_T} = \dot M_c gH_P\, \Delta X_\mathrm{melt}\, {\rho c_PT\over XP}{\chi_X\over\chi_T},
\end{equation}
where we have also used the relation $P=\rho gH_P$. 
Now applying the thermodynamic identity $\nabla_\mathrm{ad}=\chi_TP/\rho c_PT\chi_\rho$
gives
\begin{equation}
 L_H = \dot M_c gH_P\,\Delta X_\mathrm{melt}\, \left({1\over X\nabla_\mathrm{ad}}{\chi_X\over\chi_\rho}\right).
\end{equation}
This shows that the inwards convective luminosity (and compensating outwards luminosity carried by thermal conduction) is of the same order of magnitude as $L_\mathrm{chem}$.

The luminosity carried in kinetic energy on the other hand is only a small fraction of the gravitational energy release for $\tau<1$. First consider the non-rotating case. In the $\tau<1$ regime, we have $F_H\approx\rho v c_P T \nabla_\mathrm{ad}\mathrm{Pe}/C$ (using eqs.~\eqref{eq:heatflux} and \eqref{eq:nne} with $\mathrm{Pe}\ll 1$ and $\nabla\ll \nabla_\mathrm{ad}$ and setting $\ell=2H_P$ for simplicity). Comparing with the kinetic energy flux $F_K\approx (1/2)\rho v_c^3$, we find
\begin{equation}
    {F_K\over F_H} \approx {1\over 2} {v^2\over c_PT}{C\over \nabla_\mathrm{ad}\mathrm{Pe}}= {1\over 2}
    {\kappa_T^2\over H_P^2 c_PT}{C\over \nabla_\mathrm{ad}}\mathrm{Pe} =  {1\over 2}\nabla_\mathrm{ad}C{\mathrm{Pe}\over\mathrm{Ra_T}},
\end{equation} 
where we rewrite $v_c$ as $\mathrm{Pe}(\kappa_T/H_P)$ in the second step. In the rotating case, this argument can be repeated but with the substitution  $v_c\rightarrow \mathrm{Pe}(\kappa_T/L)=\mathrm{Pe}(\kappa_T/H_P)\mathrm{Ta}^{1/6}\mathrm{Pe}^{-1/3}$ instead (see equation \ref{eq:Ro}), giving
\begin{equation} \label{eq:fk_fh_low_tau}
    {F_K\over F_H} \approx  {1\over 2} \nabla_\mathrm{ad}C{(\mathrm{Pe}\mathrm{Ta})^{1/3}\over\mathrm{Ra_T}}.
\end{equation} 
With or without rotation, the ratio $F_K/F_H$ is vanishingly small for the parameters associated with white dwarf crystallization. Compositionally-driven convection is therefore very different from thermally-driven convection, where $F_H\approx \rho v_c c_P T(\nabla-\nabla_\mathrm{ad})$ and $v_c^2\approx gH_P(\chi_T/\chi_\rho)(\nabla-\nabla_\mathrm{ad})$ give $F_K/F_H\sim \nabla_\mathrm{ad}$, so that heat and kinetic energy fluxes are comparable in magnitude.

\section{The role of the kinetic energy flux at large Pe}\label{app:KE}

We showed in Appendix \ref{app:ener} that the flux of kinetic energy carried by convective motions is much smaller than the heat flux at low $\mathrm{Pe}$. However, since the heat flux saturates at large $\tau$ (where $\nabla\rightarrow\nabla_\mathrm{ad}$), we would expect the kinetic energy flux to eventually dominate as $F_X$ increases and the convection is driven more strongly. Including the kinetic energy flux, the energy balance is
\begin{equation}
    {1\over 2}\rho v_c^3 + \rho c_P \kappa_T{T\nabla\over H_P} + \rho v_c c_PT(\nabla-\nabla_e)\left({\ell\over 2H_P}\right) =0
\end{equation}
(compare eq.~[\ref{eq:energy_balance}]). Rewriting $v_c$ in terms of $\mathrm{Pe}=v_c\ell/\kappa_T$, using the definition of $\mathrm{Ra_T}$, and replacing $\nabla_e-\nabla$ in favor of $\nabla_\mathrm{ad}-\nabla$ using equation \eqref{eq:nne}, we find
\begin{equation}
    \nabla = \nabla_\mathrm{ad}\left({\mathrm{Pe}^2\over \mathrm{Pe}^2 + 2\mathrm{Pe}+2C}\right)\left[1-{\nabla_\mathrm{ad}\over\mathrm{Ra_T}}\left({H_P\over\ell}\right)^3\mathrm{Pe}(\mathrm{Pe} +C)\right],
\end{equation}
which replaces equation \eqref{eq:nablass}. This shows that the outwards kinetic energy flux is large enough to cancel the inwards heat flux when $\mathrm{Pe}^2>(\mathrm{Ra_T}/\nabla_\mathrm{ad})(\ell/H_P)^3=(8A^3/\nabla_\mathrm{ad})(H_P/\ell)$. For the example shown in Figure 1, which has $\mathrm{Ra_T}\sim 10^{10}$, this corresponds to $\mathrm{Pe}\gtrsim 10^5$ or $\tau\gtrsim 10^6$. Given the even larger values of $\mathrm{Ra_T}$ in neutron star and white dwarf applications (section \ref{sec:discussion}), we therefore do not expect the kinetic energy flux to play a significant role in energy transport.

\end{document}